\documentstyle{article}
\begin{document}
\title {
\bf \huge
Exact diagonalization of the generalized supersymmetric $t-J$ model with
boundaries
}

\author{
{\bf
Heng Fan$^a$\thanks {
Permanent address: Institute of Modern Physics, 
Northwest University, Xi'an 710069, P.R.China
}, Miki Wadati$^a$, Xiao-man Wang$^b$}\\
\normalsize
$^a$Department of Physics, Graduate School of Science,\\
\normalsize University of Tokyo, Hongo 7-3-1,\\
\normalsize Bunkyo-ku, Tokyo 113-0033, Japan.\\
\normalsize
$^b$Technology Department of Library, Northwest University, China.
}

\maketitle

\begin{abstract}
We study the generalized supersymmetric $t-J$ model with boundaries in
three different gradings: FFB, BFF and FBF.
Starting from the trigonometric R-matrix,
and in the framework of the graded quantum inverse scattering method (QISM),
we solve the eigenvalue problems for the supersymmetric $t-J$ model.
A detailed calculations are presented 
to obtain the eigenvalues and Bethe ansatz equations of the supersymmetric
$t-J$ model with boundaries in three different backgrounds.
\end{abstract}              


\newpage
\baselineskip 0.5truecm

\section{Introduction}
One-dimensional strongly correlated electron models, such as the $t-J$
model, have been attracting a great deal of interests in the context of
high-$T_c$ superconductivity.
The Hamiltonian of the $t-J$ model includes the
near-neighbour hopping ($t$) and antiferromagnetic exchange ($J$)
\cite{A,ZR}
\begin{eqnarray}
H=\sum _{j=1}^L\left\{ -t{\cal {P}}
\sum _{\sigma =\pm 1}(c_{j, \sigma}^{\dagger }
c_{j+1, \sigma }+H.c.){\cal {P}}+J({\bf S}_j{\bf S}_{j+1}
-{1\over 4}n_nn_{j+1})\right\}.
\end{eqnarray}
It is known that this model is supersymmetric and integrable
for $J=\pm 2t$ \cite{L,S}. The supersymmetric $t-J$ model was also
studied in Refs.\cite{S1,BB,S2,B0,EK},
for a review, see Ref.\cite{S3} and the references
therein.
Essler and Korepin $et~al~$ showed
that the one-dimensional Hamiltonian can be
obtained from the transfer matrix of
the two-dimensional supersymmetric exactly solvable lattice model
\cite{S2,EK}.
They used the graded QISM \cite{KIB,TF}
and obtained the eigenvalues and eigenvectors for
the supersymmetric $t-J$ model with periodic boundary
conditions in three different backgrounds, for related works, 
see for example \cite{FK}. In this paper, we shall start from the
trigonometric R-matrix which is a generalization of the R-matrix
used in \cite{EK}. The Hamiltonian is also a generalization of the
supersymmetric $t-J$ model. We shall consider the reflecting
boundary condition cases. By using the graded QISM, we obtain
the eigenvalues of the transfer matrix with boundaries in three
different backgrounds.

The exactly solvable models are generally solved by imposing 
periodic boundary conditions. Recently, solvable models with reflecting
(open) boundary conditions have been extensively studied [14-39].
Besides the original Yang-Baxter
equation \cite{Y,B}, the reflection equations also play a
key role in proving the commutativity of the transfer matrices under 
reflecting boundary conditions \cite{S4,C}.
The Hamiltonian includes non-trivial
boundary terms which are determined by the boundary K matrices.
          
In our previous paper \cite{FHS}, we used the algebraic
Bethe ansatz method to solve the eigenvalue and eigenvector problems of
the supersymmetric $t-J$ model with reflecting boundary conditions
in the framework of the graded QISM (FFB grading).
Here we shall extend the results in Ref.\cite{FHS}.
We start from
the trigonometric R-matrix proposed by Perk and Schultz \cite{PS} and change
the formulae to the graded case. Three kinds of grading are imposed, so
there are three R matrices for different grading.
Solving the graded reflection equation,
we give general diagonal solutions. There are altogether
four kinds of different boundary conditions for each choice
of grading. Using the graded algebraic
Bethe ansatz method in three possible grading
FFB, BFF and FBF, we obtain the eigenvalues of the transfer matrix
with general diagonal boundary matrices.

The graded method was proposed in \cite{OWA}, and it was applied
for the reflection equation in \cite{MN1},
and later was applied to fermionic models \cite{Z,SW}.
In this paper, we shall use the graded reflection equation to study the
supersymmetric $t-J$ model.          
For the supersymmetric $t-J$ model, the spin of the electrons and
the charge "hole" degrees of freedom play a very similar role
forming a graded superalgebra with two fermions and one boson.
The holes obey boson
commutation relations, while the spinons are fermions,see Ref.\cite{S3}
and the references therein.
The graded approach has the advantage of making a clear distinction
between bosonic and fermionic degrees of freedom. So, it is interesting
to study the supersymmetric $t-J$ model with reflecting boundary
conditions by the graded algebraic Bethe ansatz method.
In this paper, we give a detailed analysis for the Bethe ansatz
in three different backgrounds.
We should mention that the trigonometric R-matrix related to
the supersymmetric $t-J$ model with reflecting boundary conditions
was studied in \cite{FK1,G} by using the usual reflection equation, 
the results have also been extended to more general cases \cite{DG,RFH}. 
And the thermodynamic limit of the Bethe ansatz
was calculated in Ref.\cite{E}. The finite-size corrections
in the supersymmetric $t-J$ model with boundary fields are presented
in Ref.\cite{AS}.
The integrable bulk Hamiltonian was derived
previously by Karowski and Foerster and by Gonzales-Ruiz\cite{FK1,G}.
Bariev also showed that it is integrable and studied
physical properties for the Hermitian case\cite{B0}.

As mentioned in Ref.\cite{EK},
the formulae and the results for three different gradings are
significantly different, so we shall write out in detail the graded
algebraic Bethe ansatz for the generalized supersymmetric $t-J$ model
with four kinds of boundaries.

The paper is organized as follows: In section 2,
we review the supersymmetric $t-J$ model and its generalization.
We start from
the Perk-Shultz\cite{PS} model
and change it to the graded case. In section 3,
the general solutions of the reflection equation
are presented. In section 4, in the FFB grading, we use the algebraic
Bethe ansatz method to obtain the eigenvalues and eigenvectors of
the transfer matrix with boundaries. In sections 5 and 6, we study
the case of BFF grading and FBF grading.
Section 7 includes a brief summary and
some discussions. 
                                             
\section{Supersymmetric $t-J$ model and its generalization}
We first review the supersymmetric $t-J$ model. For
convenience, we adopt the notations in Ref.\cite{EK}.
The Hamiltonian of the supersymmetric $t-J$ model is given as:
\begin{eqnarray}
H&=&-t\sum _{j=1}^N\sum _{\sigma =\pm }
[c_{j,\sigma }^{\dagger}(1-n_{j,-\sigma })c_{j+1,\sigma }
(1-n_{j+1,-\sigma})
+c_{j+1,\sigma }^{\dagger}(1-n_{j+1,-\sigma })c_{j+1,\sigma }
(1-n_{j,-\sigma})]
\nonumber \\
&&+J\sum _{j=1}^N[S_j^zS_{j+1}^z+{1\over 2}(S_j^{\dagger }S_{j+1}
+S_jS_{j+1}^{\dagger })-{1\over 4}n_jn_{j+1}].
\end{eqnarray}
This form is an equivalent expression of the Hamiltonian (1).
The operators $c_{j,\sigma }$ and $c_{j,\sigma }^{\dagger }$ mean the
annihilation and creation operators of electron with spin $\sigma $
on a lattice site $j$, and we assume the total number of
lattice sites is $N$, $\sigma =\pm $ represent
spin down and up, respectively. These operators are canonical
Fermi operators satisfying anticommutation relations
\begin{eqnarray}
\{c_{j,\sigma }^{\dagger },c_{j,\tau }\} =\delta _{ij}
\delta _{\sigma \tau }.
\end{eqnarray}
We denote by $n_{j,\sigma }=c_{j,\sigma }^{\dagger }c_{j,\sigma }$
the number operator for the electron on a site $j$ with
spin $\sigma $, and by
$n_j=\sum _{\sigma =\pm }n_{j,\sigma }$ the number operator for the
electron on a site $j$.
The Fock vacuum state $|0>$ satisfies $c_{j,\sigma }|0>=0$. There are
altogether three possible electronic states at a given lattice site
$j$ due to excluding double occupancy:
\begin{eqnarray}
|0>,~~~|\uparrow >_j=c_{j,1}^{\dagger }|0>,~~~
|\downarrow >_j=c_{j,-1}^{\dagger }|0>.
\end{eqnarray}
$S^z_j,S_j,S_j^{\dagger }$ are spin operators satisfying $su(2)$
algebra and can be expressed as:
\begin{eqnarray}
S_j=c_{j,1}^{\dagger }c_{j,-1},
~~~~S_j^{\dagger }=c_{j,-1}^{\dagger }c_{j,},
~~~~S_j^z={1\over 2}(n_{j,1}-n_{j,-1}).
\label{S}
\end{eqnarray}

It has been proved that for a special value
$J=2t=2$, the Hamiltonian of the supersymmetric $t-J$ model
can be written as the a graded permutation operator \cite {BB,S2,EK}
\begin{eqnarray}
H=-\sum _{j=1}^NP_{j,j+1}-2\hat {N}.
\end{eqnarray}
Here we have omitted a constant term. The total number operator
$\hat {N}=\sum _{j=1}^Nn_j$ commutes with the Hamiltonian and
is dedicated to the chemical potential. We shall also omit
the second term in the following. The graded permutation
operator can be represented as
\begin{eqnarray}
P_{ac}^{bd}=\delta _{ad}\delta _{bc}(-1)^{\epsilon _a\epsilon _c}.
\end{eqnarray}
Here, different from the non-graded case,
we have the Grassmann parities $\epsilon _a=1,0$ representing
fermion and boson, respectively. The Hamiltonian can also be
represented by the generators of $u(1|2)$, $su(1|2)$ is a subalgebra of
$u(1|2)$,
\begin{eqnarray}
H=-\sum _{j=1}^N
\left\{ \sum _{\sigma }
(Q_{j+1,\sigma }^{\dagger }Q_{j,\sigma }
+Q_{j,\sigma }^{\dagger }Q_{j+1,\sigma })
-2S_j^zS_{j+1}^z-S_jS_{j+1}^{\dagger }-S_{j+1}S_j^{\dagger }
+2T_jT_{j+1}\right\}.
\end{eqnarray}
The generators of the algebra $u(1|2)$ are given by relation (5) and the
following:
\begin{eqnarray}
Q_{j,\pm }=(1-n_{j,\mp })c_{j,\pm },
~~~~Q_{j,\pm }^{\dagger }=(1-n_{j,\mp })c_{j,\pm }^{\dagger },
~~~~T_j=1-{1\over 2}n_j.
\end{eqnarray}
The fundamental representations of these operators take the following form
\begin{eqnarray}
S_j^z=\left[ \begin{array}{ccc}
-{1\over 2}&0&0\\ 0&{1\over 2}&0\\ 0&0&0 \end{array}\right],
T_j=\left[ \begin{array}{ccc}
{1\over 2}&0&0\\ 0&{1\over 2}&0\\ 0&0&1 \end{array}\right],
\nonumber \\
S_k=e^k_{21}, ~~~S_k^{\dagger }=e^k_{12},
~~~~Q_{k,1}=e^k_{32},
\nonumber \\
Q_{k,1}^{\dagger }=e^k_{23},
~~~~Q_{k,-1}=e^k_{31},~~~~Q_{k,-1}^{\dagger }=e^k_{13},
\end{eqnarray}
where $e_{ij}^k$ is a $3\times 3$ matrix acting on the $k$-th space with
elements $(e_{ij}^k)_{\alpha \beta }=\delta _{i\alpha }
\delta _{j\beta }$.

The above Hamiltonian can be obtained from the logarithmic derivative at
zero spectral parameter of the transfer matrix
constructed by the rational R-matrix.
In this paper, we shall study the trigonometric R-matrix.
Let's start from the R-matrix of the Perk-Schultz
model \cite{PS}, the non-zero
entries of the R-matrix are given by
\begin{eqnarray}
&&{\tilde {R}}(\lambda )^{aa}_{aa}=sin(\eta +\epsilon _a\lambda ),
\nonumber \\
&&{\tilde {R}}(\lambda )^{ab}_{ab}=q_{ab}
sin(\lambda ), ~~a\not= b,
\nonumber \\
&&{\tilde {R}}(\lambda )^{ab}_{ba}=e^{isign(a-b)\lambda }sin(\eta ),
~~a\not= b,
\end{eqnarray}
where
\begin{eqnarray}
sign(a-b)=\left\{
\begin{array}{ll}1,&{\rm if} ~a>b \\
-1, &{\rm if}~ a<b.
\end{array}\right.
\end{eqnarray}
As mentioned above, $\epsilon _a$ is the Grassman parity,
$\epsilon _a=0$ for boson and $\epsilon _a=1$ for fermion.
We demand $q_{ab}q_{ba}=1$, in the following, and
let $q_{ab}=(-)^{\epsilon _a\epsilon _b}$.
This R-matrix of the Perk-Schultz model                
satisfies the usual Yang-Baxter equation:
\begin{eqnarray}
{\tilde {R}}_{12}(\lambda -\mu )
{\tilde {R}}_{13}(\lambda )
{\tilde {R}}_{23}(\mu )
={\tilde {R}}_{23}(\mu )
{\tilde {R}}_{13}(\lambda )                                    
{\tilde {R}}_{12}(\lambda -\mu )
\end{eqnarray}
Introducing a diagonal matrix
$I_{ac}^{bd}=(-)^{\epsilon _a\epsilon _c}
\delta _{ab}\delta _{cd}$, we change the original R-matrix to the following
form
\begin{eqnarray}
R(\lambda )=I{\tilde {R}}(\lambda ).
\end{eqnarray}
Considering the non-zero elements of the R-matrix
$R_{ab}^{cd}$, we have
$\epsilon _a+\epsilon _b+\epsilon _c+\epsilon _d=0$.
One can show that the R-matrix satisfies the graded Yang-Baxter equation
\begin{eqnarray}
R(\lambda -\mu )_{a_1a_2}^{b_1b_2}
R(\lambda )_{b_1a_3}^{c_1b_3}
R(\mu )_{b_2b_3}^{c_2c_3}
(-)^{(\epsilon _{b_1}+\epsilon _{c_1})\epsilon _{b_2}}
=
R(\mu )_{a_2a_3}^{b_2b_3}R(\lambda )_{a_1b_3}^{b_1c_3}
R(\lambda -\mu )_{b_1b_2}^{c_1c_2}(-)^{(\epsilon _{a_1}
+\epsilon _{b_1})\epsilon _{b_2}}.
\end{eqnarray}
In the framework of the QISM, we can construct 
the $L$ operator from the R-matrix as:
\begin{eqnarray}
L_{aq}(\lambda )\equiv R_{aq}(\lambda ),
\end{eqnarray}
where $a$ represents the auxiliary space and $q$
represents the quantum space.
Thus we have the (graded) Yang-Baxter relation
\begin{eqnarray}
R_{12}(\lambda -\mu )L_1(\lambda )L_2(\mu )
=L_2(\mu )L_1(\lambda )R_{12}(\lambda -\mu ).
\end{eqnarray}
Here the tensor product is in the sense of super tensor product
defined as
\begin{eqnarray}
(F\otimes G)_{ac}^{bd}=F_a^bG_c^d(-)^{(\epsilon _a+\epsilon _b)\epsilon _c}.
\end{eqnarray}
In the rest of this paper, all tensor products are in the super sense.
However, there are two kinds of super tensor product, we shall point it
out later.

The row-to-row monodromy matrix
$T_N(\lambda )$ is defined as the matrix product over
the $N$ operators on all sites of the lattice, 
\begin{eqnarray}
T_a(\lambda )=L_{aN}(\lambda )L_{aN-1}(\lambda )\cdots L_{a1}(\lambda ), 
\end{eqnarray}
where $a$ still represents the auxiliary space, and
the tensor product is in the graded sense. Explicitely we write
\begin{eqnarray}
&&\{ [T(\lambda )]^{ab}\}_{\begin {array}{c}
\alpha _1\cdots \alpha _N\\
\beta _1\cdots \beta _N\end{array}}
\nonumber \\
&=&L_N(\lambda )_{a\alpha _N}^{c_N\beta _N}
L_{N-1}(\lambda )_{c_N\alpha _{N-1}}^{c_{N-1}\beta _{N-1}}
\cdots L_1(\lambda )_{c_2\alpha _1}^{b\beta _1}
(-1)^{\sum _{j=2}^N(\epsilon _{\alpha _j}+\epsilon _{\beta _j})
\sum _{i=1}^{j-1}\epsilon _{\alpha _i}}
\end{eqnarray}
By repeatedly using the Yang-Baxter relation (17), one can prove easily that
the monodromy matrix also satisfies the Yang-Baxter relation
\begin{eqnarray}
R(\lambda -\mu )T_1(\lambda )T_2(\mu )
=T_2(\mu )T_1(\lambda )R(\lambda -\mu ).
\end{eqnarray}

For periodic boundary condition, the transfer
matrix $\tau _{peri}(\lambda )$
of this model is defined as the supertrace of the
monodromy matrix in the auxiliary space. In general case,
the supertrace is defined as
\begin{eqnarray}
\tau _{peri}(\lambda )=strT(\lambda )
=\sum (-1)^{\epsilon _a}T(\lambda )_{aa}.
\end{eqnarray}

As a consequence of the Yang-Baxter relation (21) and the unitarity
property of the R-matrix, we can prove that
the transfer matrix commutes with each other for different spectral
parameters.
\begin{eqnarray}
[\tau _{peri}(\lambda ),\tau _{peri}(\mu )]=0 
\end{eqnarray}
Generally in this sense we mean the model is
integrable. Expanding the transfer matrix in the powers of $\lambda $,
we can find conserved quantites, the first non-trivial conserved
equantity is the Hamiltonian.

For the rational R-matrix,
it has been proved that the Hamiltonian obtained by taking the first
logarithmic derivative at the zero spectral parameter, 
$H=-i\frac {d\ln [\tau (\lambda )]}{d\lambda }|_{\lambda =0}
=-\sum _{k=1}^NP_{k,k+1}$,
is equivalent to the Hamiltonian of the supersymmtric $t-J$ model \cite{EK}.

Here we shall study the trigonometric case. Noting
$R_{ij}(0)=-sin(\eta )P_{ij}$, the Hamiltonian can be defined as:
\begin{eqnarray}
H=sin(\eta )
\frac {d\ln [\tau (\lambda )]}{d\lambda }|_{\lambda =0}
=\sum _{j=1}^NH_{j,j+1},
\end{eqnarray}
with $H_{j,j+1}\equiv P_{j,j+1}L'_{j,j+1}(0)$.

As an example,
we choose Fermionic, Fermionic and Bosonic (FFB) grading that means
$\epsilon _1=\epsilon _2=1, \epsilon _3=0$.
Explicitly, we can write the R-matrix as:
\begin{eqnarray}
R(\lambda )=\left(
\begin{array}{ccccccccc}
a(\lambda )&0&0&0&0&0&0&0&0\\                                        
0&b(\lambda )&0&-c_-(\lambda )&0&0&0&0&0\\
0&0&b(\lambda )&0&0&0&c_-(\lambda )&0&0\\
0&-c_+(\lambda )&0&b(\lambda )&0&0&0&0&0\\
0&0&0&0&a(\lambda )&0&0&0&0\\
0&0&0&0&0&b(\lambda )&0&c_-(\lambda )&0\\
0&0&c_+(\lambda )&0&0&0&b(\lambda )&0&0\\
0&0&0&0&0&c_+(\lambda )&0&b(\lambda )&0\\
0&0&0&0&0&0&0&0&w(\lambda )
\end{array}\right),
\end{eqnarray}
where
\begin{eqnarray}
a(\lambda )=sin(\lambda -\eta ),
~w(\lambda )=sin(\lambda +\eta ), 
~b(\lambda )=sin(\lambda ),
~c_{\pm }(\lambda )=e^{\pm i\lambda }sin(\eta ).
\end{eqnarray}
The rational limit of this R-matrix is completely the same as the one used
by Essler and Korepin in Ref.\cite{EK}.
In the framework of the QISM, we define
the $L$ operator as
\begin{eqnarray}
L_n(\lambda )=\left(
\begin{array}{ccc}
b(\lambda )-(b(\lambda )-a(\lambda ))e^n_{11}&
-c_-(\lambda )e^n_{21} &c_-(\lambda )e^n_{31}\\
-c_+(\lambda )e^n_{12} &b(\lambda )-(b(\lambda )-a(\lambda ))e^n_{22}&
c_-(\lambda )e^n_{32}\\
c_+(\lambda )e^n_{13} &c_+(\lambda )e^n_{23} &
b(\lambda )-(b(\lambda )-w(\lambda ))e^n_{33}
\end{array}\right).                                         
\end{eqnarray}
Here $e^n_{ab}$ acts on the $n$-th quantum space.

We denote explicitly the row-to-row monodromy matrix as
\begin{eqnarray}
T(\lambda )=
\left( \begin{array}{ccc}
A_{11}(\lambda )&A_{12}(\lambda ) &B_1(\lambda )\\
A_{21}(\lambda )&A_{22}(\lambda )&B_2(\lambda )\\
C_1(\lambda )&C_2(\lambda )&D(\lambda )\end{array}
\right).
\end{eqnarray}
If we choose the FFB grading,
the transfer matrix is then given as
\begin{eqnarray}
\tau (\lambda )_{peri}=-A_{11}(\lambda )-A_{22}(\lambda )+D(\lambda ).
\end{eqnarray}
Thus we can write
\begin{eqnarray}
L'(0)=\left(
\begin{array}{ccc}
1-(1-cos(\eta ))e_{11} &isin(\eta )e_{21} &-isin(\eta )e_{31}\\
-isin(\eta )e_{12}&1-(1-cos(\eta ))e_{22}&-isin(\eta )e_{32}\\
isin(\eta )e_{13}&isin(\eta )e_{23} &1-(1-cos(\eta ))e_{33}
\end{array}\right).
\end{eqnarray}
With the help of the fundamental representation of algebra $u(1|2)$,
we have
\begin{eqnarray}
H_{j,j+1}
&=&\sum _{\sigma =\pm }[Q_{j,\sigma }Q_{j+1,\sigma }^{\dagger }
+Q_{j,\sigma }^{\dagger }Q_{j+1,\sigma }]
-S_jS_{j+1}^{\dagger }-S_{j+1}S_j^{\dagger }
+con(\eta )[-2S_j^zS_{j+1}^z+2T_jT_{j+1}^{\dagger }]
\nonumber \\
&&+2isin(\eta )[-S_j^zT_{j+1}+T_jS_{j+1}^z+S_j^z-S_{j+1}^z+T_j-T_{j+1}].
\end{eqnarray}                                               
As mentioned in introduction, this Hamiltonian was previously obtained
by Karowski and Foerster and by Gonzales-Ruiz\cite{FK1,G}.
Explicitly, using the fermionic representation (5) and (9),
we can write the Hamiltonian of the generalized supersymmetric
$t-J$ model as following\cite{FK1, G}:
\begin{eqnarray}
H&=&\sum _{j=1}^N\sum _{\sigma =\pm }
[c_{j,\sigma }^{\dagger}(1-n_{j,-\sigma })c_{j+1,\sigma }
(1-n_{j+1,-\sigma})
+c_{j+1,\sigma }^{\dagger}(1-n_{j+1,-\sigma })c_{j+1,\sigma }
(1-n_{j,-\sigma})]
\nonumber \\
&&-2\sum _{j=1}^N[{1\over 2}(S_j^{\dagger }S_{j+1}+S_jS_{j+1}^{\dagger })
+cos(\eta )S_j^zS_{j+1}^z
-{cos(\eta )\over 4}n_jn_{j+1}]
\nonumber \\
&&+isin(\eta )\sum _{j=1}^N[S_j^zn_{j+1}-S_{j+1}^zn_j].
\end{eqnarray}
Here periodic boundary condition is assumed.
We remark that this Hamiltonian is in general not Hermitian.

In this paper, we shall study the reflecting boundary conditions,
which may cause non-trivial boundary terms in the Hamiltonian.

\section{Integrable reflecting boundary conditions and the
solutions of reflection equation}
In this paper, we consider the reflecting boundary condition case.
In the end of 80's,
Sklyanin proposed a systematic approach to handle the exactly
solvable models with reflecting (open) boundary conditions\cite{S4}, which
includes a so-called reflection equation proposed by Cherednik\cite{C}.
\begin{eqnarray}
R_{12}(\lambda -\mu )K_1(\lambda )R_{21}(\lambda +\mu )
K_2(\mu )=K_2(\mu )R_{12}(\lambda +\mu )K_1(\lambda )         
R_{21}(\lambda -\mu )
\end{eqnarray}
For the graded case, the above form of the reflection equation remains the
same. We only need to change the usual tensor product to the graded
tensor product \cite{MN1}. We write it explicitly as
\begin{eqnarray}
&&R(\lambda -\mu )_{a_1a_2}^{b_1b_2}K(\lambda )_{b_1}^{c_1}
R(\lambda +\mu )_{b_2c_1}^{c_2d_1}                           
K(\mu )_{c_2}^{d_2}(-)^{(\epsilon _{b_1}+\epsilon _{c_1})\epsilon _{b_2}}
\nonumber \\
&=&K(\mu )_{a_2}^{b_2}R(\lambda +\mu )_{a_1b_2}^{b_1c_2}     
K(\lambda )_{b_1}^{c_1}
R(\lambda -\mu )_{c_2c_1}^{d_2d_1}                            
(-)^{(\epsilon _{b_1}+\epsilon _{c_1})\epsilon _{c_2}}.
\end{eqnarray}

We concentrate the discussion to the diagonal solutions of the reflection
equation. Suppose $K(\lambda )_{a}^{b}=\delta _{ab}k_a(\lambda )$.
Inserting this relation into the reflection equation, we find there
are only one non-trivial relation to be solved:
\begin{eqnarray}
&&R(\lambda -\mu )_{a_1a_2}^{a_1a_2}R(\lambda +\mu )_{a_2a_1}^{a_1a_2}
k(\lambda )_{a_1}k(\mu )_{a_1}
+R(\lambda -\mu )_{a_1a_2}^{a_2a_1}
R(\lambda +\mu )_{a_1a_2}^{a_1a_2}k(\lambda )_{a_2}
k(\mu )_{a_1}
\nonumber \\
&=&R(\lambda +\mu )_{a_1a_2}^{a_1a_2}R(\lambda -\mu )_{a_2a_1}^{a_1a_2}
k(\mu )_{a_2}k(\lambda )_{a_1}
+R(\lambda +\mu )_{a_1a_2}^{a_2a_1}                          
R(\lambda -\mu )_{a_1a_2}^{a_1a_2}k(\mu )_{a_2}k(\lambda )_{a_2}
\end{eqnarray}
Suppose $a_2>a_1$, and substitute the exact form of the elements of
R-matrix into the above relation.
We find a general diagonal solution:
\begin{eqnarray}               
\frac {k(\lambda )_{a_1}}{k(\lambda )_{a_2}}
=\frac {sin(\xi +\lambda )}{sin(\xi -\lambda )}e^{-2i\lambda },
\end{eqnarray}
where $\xi $ is an arbitrary parameter. In a special limit we can see
the identity is also a solution of the reflection equation.
For the cases (FFB, BFF and FBF grading) we study in this paper,
there are two types of solutions to
the reflection equation
\begin{eqnarray}
K_{I}(\lambda )&=&\left(
\begin{array}{ccc}
sin(\xi +\lambda )e^{-2i\lambda }&&\\
&sin(\xi +\lambda )e^{-2i\lambda }&\\
&&sin(\xi -\lambda )\end{array}\right),                    
\nonumber \\
K_{II}(\lambda )&=&\left(
\begin{array}{ccc}
sin(\xi +\lambda )e^{-2i\lambda }&&\\
&sin(\xi -\lambda )&\\
&&sin(\xi -\lambda )\end{array}\right).
\end{eqnarray}

Instead of the monodromy matrix $T(\lambda )$ for periodic boundary
conditions, we consider the double-row monodromy matrix
\begin{eqnarray}
{\cal {T}}(\lambda )=T(\lambda )K(\lambda )T^{-1}(-\lambda )
\end{eqnarray}
for the reflecting boundary conditions. Using the Yang-Baxter relation, and
considering the boundary K-matrix which satisfies
the reflection equation, one
can prove that the double-row monodromy matrix ${\cal {T}}(\lambda )$
also satisfies the reflection equation
\begin{eqnarray}
&&R(\lambda -\mu )_{a_1a_2}^{b_1b_2}{\cal {T}}(\lambda )_{b_1}^{c_1}
R(\lambda +\mu )_{b_2c_1}^{c_2d_1}                           
{\cal {T}}(\mu )_{c_2}^{d_2}
(-)^{(\epsilon _{b_1}+\epsilon _{c_1})\epsilon _{b_2}}
\nonumber \\
&=&{\cal {T}}(\mu )_{a_2}^{b_2}R(\lambda +\mu )_{a_1b_2}^{b_1c_2}     
{\cal {T}}(\lambda )_{b_1}^{c_1}
R(\lambda -\mu )_{c_2c_1}^{d_2d_1}                            
(-)^{(\epsilon _{b_1}+\epsilon _{c_1})\epsilon _{c_2}}.
\end{eqnarray}

Next, we shall study the properties of the R-matrix.
We define the super-transposition $st$ as
\begin{eqnarray}
(A^{st})_{ij}=A_{ji}(-1)^{(\epsilon _i+1)\epsilon _j}.
\end{eqnarray}
As an example, we take the FFB grading,
that means $\epsilon _1=\epsilon _2=1
, \epsilon _3=0$. We can rewrite the above relation explicitly as
\begin{eqnarray}
\left( \begin{array}{ccc}
A_{11}&A_{12}&B_1\\
A_{21}&A_{22}&B_2\\
C_1&C_2&D\end{array}\right)^{st}                            
=\left( \begin{array}{ccc}
A_{11}&A_{21}&C_1\\
A_{12}&A_{22}&C_2\\
-B_1&-B_2&D\end{array}\right).
\end{eqnarray}
We also define the inverse of 
the super-transposition $\bar {st}$ as 
$\{ A^{st}\} ^{\bar {st}}=A$.

For the R-matrix with all three different grading,
FFB, BFF and FBF, we can prove directly that
the R-matrix satisfy the following unitarity and cross-unitarity
relations:
\begin{eqnarray}
&&R_{12}(\lambda )R_{21}(-\lambda )=\rho (\lambda )\cdot id.,~~~~
\rho (\lambda )=sin(\eta +\lambda )sin(\eta -\lambda ),    \\
&&R_{12}^{st_1}(\eta -\lambda )M_1
R_{21}^{st_1}(\lambda )M_1^{-1}={\tilde {\rho }}(\lambda )\cdot id.,~~~~
\tilde {\rho }(\lambda )=sin(\lambda )sin(\eta -\lambda ).
\end{eqnarray}
Here the matrix $M$ is diagonal and is determined by the R-matrix.
For three different gradings, the forms of $M$ are different. 
We have: $M=diag.(e^{2i\eta },1,1)$ for FFB grading,
$M=diag.(1,1, e^{-2i\eta })$ for BFF grading and $M=1$ for FBF grading.

In order to construct the commuting transfer matrix with boundaries,
besides the
reflection equation, we need the dual reflection equation.
Generally, the dual reflection equation which depends on
the unitarity and cross-unitrarity relations of the R-matrix   
takes different forms for different models. 
For the models considered in this paper, the cross-unitarity relation
remains the same for three different back backgrounds. We can write the dual
reflection equation in the following form:
\begin{eqnarray}
&&R_{12}(\mu -\lambda )K_1^+(\lambda )M_1^{-1}R_{21}(\eta -\lambda -\mu )
K_2^+(\mu )M_2^{-1}     
\nonumber \\
&=&K_2^+(\mu )M_2^{-1}R_{12}(\eta -\lambda -\mu )
K_1^+(\lambda )M_1^{-1}R_{21}(\mu -\lambda ).
\end{eqnarray}
One finds that there is an isomorphism between the reflection 
equation (33) and the dual reflection equation (44)
\begin{eqnarray}
K(\lambda ):\rightarrow
K^+(\lambda )=MK(-\lambda +\eta /2).
\end{eqnarray}
Here we mean: given a solution of
the reflection equation (33), we can find a solution of the 
dual reflection equation (44). Note, however,
that in the sense of the commuting
transfer matrix, the reflection equation and the dual reflection equation
are independent of each other.

The transfer matrix with boundaries is defined as:
\begin{eqnarray}
t(\lambda )=strK^+(\lambda ){\cal {T}}(\lambda ).
\end{eqnarray}
The commutativity of $t(\lambda )$ can be proved by using
unitarity and cross-unitarity relations, reflection equation and 
the dual reflection equation. The detailed proof of the commuting
transfer matrix with boundaries for super (graded)
case can be found, for instance, in Ref.\cite{BGZZ,GZZ,FHS,FG} etc..

We also define the Hamiltonian by a relation                        
\begin{eqnarray}
H&\equiv &{1\over 2}sin(\eta )\frac {d\ln t(\lambda )}
{d\lambda}|_{\lambda =0}
\nonumber \\
&=&\sum _{j=1}^{N-1}P_{j,j+1}L'_{j,j+1}(0)+
{1\over 2}\frac {sin(\eta )}{sin(\xi )}K'_1(0)
+\frac {str_aK_a^+(0)P_{Na}L_{Na}'(0)}{str_{a}K_a^+(0)}.
\end{eqnarray}
We still take the FFB grading as an example,
and thus $M=diag.(e^{2i\eta },1,1)$.
We have two types of the solutions to 
the dual reflection equation
\begin{eqnarray}
K^+_{I}(\lambda )&=&\left(
\begin{array}{ccc}
sin(\xi ^+-\lambda )e^{i(2\lambda +\eta )}&&\\
&sin(\xi ^+-\lambda )e^{i(2\lambda -\eta)}&\\
&&sin(\xi ^++\lambda -\eta )\end{array}\right),                    
\nonumber \\
K^+_{II}(\lambda )&=&\left(
\begin{array}{ccc}
sin(\xi ^+-\lambda )e^{i(2\lambda +\eta )}&&\\
&sin(\xi ^++\lambda -\eta )&\\
&&sin(\xi ^++\lambda -\eta )\end{array}\right),
\end{eqnarray}
where $\xi ^+$ is also an arbitrary boundary parameter.
Since the reflection equation and the dual reflection equation are
independent of each other, there are
altogether four different types of boundaries determined by boundary
$K$ and $K^+$ matrices: $\{K_I,K_I^+\}$, $\{K_I,K_{II}^+\}$,
$\{K_{II},K_I^+\}$, $\{K_{II},K_{II}^+\}$.

The Hamiltonian of the generalized
supersymmetric $t-J$ model with boundaries is written as
\begin{eqnarray}
H&=&\sum _{j=1}^{N-1}\sum _{\sigma =\pm }
[c_{j,\sigma }^{\dagger}(1-n_{j,-\sigma })c_{j+1,\sigma }
(1-n_{j+1,-\sigma})
+c_{j+1,\sigma }^{\dagger}(1-n_{j+1,-\sigma })c_{j+1,\sigma }
(1-n_{j,-\sigma})]
\nonumber \\
&&-2\sum _{j=1}^{N-1}[{1\over 2}(S_j^{\dagger }S_{j+1}+S_jS_{j+1}^{\dagger })
+cos(\eta )S_j^zS_{j+1}^z
-{cos(\eta )\over 4}n_jn_{j+1}]
\nonumber \\
&&+isin(\eta )\sum _{j=1}^{N-1}[S_j^zn_{j+1}-S_{j+1}^zn_j]
-2cos(\eta )\sum _{j=1}^{N-1}n_j+e^{-i\eta }n_1-e^{-i\eta }n_N
+H_1+H_{N},
\end{eqnarray}
where $H_1$ and $H_N$ are determined by the reflecting matrices.
Explicitly, they are
\begin{eqnarray}
&&H_1^I=\frac {sin(\eta )}{sin(\xi )}e^{i\xi }n_1,
~~~~H_1^{II}=\frac {sin(\eta )}{2sin(\xi )}e^{i\xi }n_1
-\frac {sin(\eta )}{sin(\xi )}e^{i\xi }S_1^z,
\nonumber \\
&&H_N^I=-\frac {sin(\eta )}{2sin(\xi ^++\eta )}e^{-i(\xi ^++\eta )}n_N,
~~~~H_N^{II}=-\frac {sin(\eta )}{2sin(\xi ^+)}e^{-i\xi ^+}n_N
+\frac {sin(\eta )}{sin(\xi ^+)}e^{-i\xi ^+}S_N^z.
\end{eqnarray}
We remark that there are four types of boundary terms in the Hamiltonian.

The solution of the graded reflection equation is identical to that of the
non-graded reflection equation, because we focus our attention
on the diagonal
solutions of the reflectione equation, and the two cases for graded
and non-graded is completely the same.
The solution of the dual reflection equation for FFB case is
similar to the non-graded case in Ref.\cite{G} except a minus in the
last diagonal elements.
And the boundary terms appeared
in the Hamiltonian (49,50) are similar to the previous results\cite{G} 
(the anisotropic parameter should be redefined as $\eta \equiv -\gamma $).

\section{Algebraic Bethe ansatz method for FFB grading}
In this section, the FFB grading is assumed.
We shall use the nested algebraic Bethe ansatz
method to obtain
the eigenvalues of the transfer matrix with boundaries defined above.

\subsection{Commutation relations necessary for the algebraic
Bethe ansatz method}

We write solution of the dual reflection equation $K^+$ and the
double-row monodromy matrix ${\cal {T}}$ respectively
in the following form:
\begin{eqnarray}
K^+(\lambda )&=&diag.(k_1^+(\lambda ), k_2^+(\lambda ),
k_3^+(\lambda )),
\\
{\cal {T}}(\lambda )
&=&\left( \begin{array}{ccc}
{\cal {A}}_{11}(\lambda )&{\cal {A}}_{12}(\lambda ) &
{\cal {B}}_1(\lambda )\\
{\cal {A}}_{21}(\lambda )&{\cal {A}}_{22}(\lambda )
&{\cal {B}}_2(\lambda )\\
{\cal {C}}_1(\lambda )&{\cal {C}}_2(\lambda )&{\cal {D}}
(\lambda )\end{array}                                          
\right) .
\end{eqnarray}
                                                               
Instead of ${\cal {A}}_{ab}$, we shall use
$\tilde {\cal {A}}_{ab}$ in the algebraic Bethe ansatz method
so that there will exist only one type wanted terms in the
commutation relation. The transformation takes the form
\begin{eqnarray}
{\cal {A}}(\lambda )_{ab}=\tilde {\cal {A}}(\lambda )_{ab}
+\delta _{ab}
\frac {e^{-2i\lambda }sin(\eta )}{sin(2\lambda +\eta )}
{\cal {D}}(\lambda ).
\end{eqnarray}
So, the transfer matrix with boundaries can be rewrite as
\begin{eqnarray}
t(\lambda )&=&-k_1^+(\lambda ){\cal {A}}_{11}(\lambda )
-k_2^+(\lambda ){\cal {A}}_{22}(\lambda )                        
+k_3^+(\lambda ){\cal {D}}(\lambda )
\nonumber \\
&=&-k_1^+(\lambda )\tilde {\cal {A}}_{11}(\lambda )
-k_2^+(\lambda )\tilde {\cal {A}}_{22}(\lambda )                
+U_3^+(\lambda )
{\cal {D}}(\lambda ),
\end{eqnarray}
where 
\begin{eqnarray}
U_3^+(\lambda )\equiv
k_3^+(\lambda )-
\frac {e^{-2i\lambda }sin(\eta )}{sin(2\lambda +\eta )}
(k_1^+(\lambda )+k_2^+(\lambda )).
\end{eqnarray}
For type I, II solutions of the dual reflection equation $K^+$, we have
\begin{eqnarray}
U_3^+(\lambda )&=&\frac {sin(2\lambda -\eta )sin(\xi ^++\lambda +\eta )}
{sin(2\lambda +\eta )}, ~~~{\rm for}~~K_I^+,
\\
U_3^+(\lambda )&=&\frac {sin(2\lambda -\eta )sin(\xi ^++\lambda )e^{i\eta }}
{sin(2\lambda +\eta )}, ~~~{\rm for}~~K_{II}^+.
\end{eqnarray}

As mentioned above, the double-row monodromy matrix also satisfies the
graded reflection equation (39). Setting the indices in that relation
to be special values, we can find the
commutation relations which are necessary for the algebraic
Bethe ansatz method. The detailed calculation is tedious and
complicated, so we do not present it here. The result is

\begin{eqnarray}
{\cal {C}}_{d_1}(\lambda ){\cal {C}}_{d_2}(\mu )
&=&-\frac {r_{12}(\lambda -\mu )_{c_2c_1}^{d_2d_1}}{sin(\lambda -\mu +\eta )}
{\cal {C}}_{c_2}(\mu ){\cal {C}}_{c_1}(\lambda ),
\end{eqnarray}

\begin{eqnarray}
{\cal {D}}(\lambda ){\cal {C}}_{d}(\mu )
&=&\frac {sin(\lambda +\mu )sin(\lambda -\mu -\eta )}
{sin(\lambda +\mu +\eta )sin(\lambda -\mu )}
{\cal {C}}_{d}(\mu )
{\cal {D}}(\lambda )
\nonumber \\
&+&
\frac {sin(2\mu )sin(\eta )e^{i(\lambda -\mu )}}
{sin(\lambda -\mu )sin(2\mu +\eta )}  
{\cal {C}}_{d}(\lambda )
{\cal {D}}(\mu )
-\frac {sin(\eta )e^{i(\lambda +\mu )}}
{sin(\lambda +\mu +\eta )}
{\cal {C}}_b(\lambda )\tilde {\cal {A}}_{bd}(\mu ),
\end{eqnarray}

\begin{eqnarray}
\tilde {\cal {A}}_{a_1d_1}(\lambda ){\cal {C}}_{d_2}(\mu )
&=&\frac {r_{12}(\lambda +\mu +\eta )_{a_1c_2}^{c_1b_2}
r_{21}(\lambda -\mu )_{b_1b_2}^{d_1d_2}}
{sin(\lambda +\mu +\eta )sin(\lambda -\mu )}                
{\cal {C}}_{c_2}(\mu )
\tilde {\cal {A}}_{c_1b_1}(\lambda )
\nonumber \\
&+&
\frac {sin(\eta )e^{-i(\lambda -\mu )}}
{sin(\lambda -\mu )sin(2\lambda +\eta )}  
r_{12}(2\lambda +\eta )_{a_1b_1}^{b_2d_1}
{\cal {C}}_{b_1}(\lambda )
\tilde {\cal {A}}_{b_2d_2}(\mu )
\nonumber \\
&-&\frac {sin(2\mu )sin(\eta )e^{-i(\lambda +\mu )}}
{sin(\lambda +\mu +\eta )sin(2\lambda +\eta )sin(2\mu +\eta )}
r_{12}(2\lambda +\eta )_{a_1b_2}^{d_2d_1}
{\cal {C}}_{b_2}(\lambda ){\cal {D}}(\mu ).
\end{eqnarray}
Here the indices take values 1,2, and the r-matrix is defined as
\begin{eqnarray}
r_{12}(\lambda )=\left( \begin{array}{cccc}
sin(\lambda -\eta )&0&0&0\\
0&sin(\lambda )&-sin(\eta )e^{-i\lambda }&0\\
0&-sin(\eta )e^{i\lambda }&sin(\lambda )&0\\
0&0&0&sin(\lambda -\eta )\end{array}\right) 
\end{eqnarray}
In fact, the elements of the r-matrix are equal to
those of the original R-matrix when its indices just take values 1,2.

\subsection{Vacuum State} 
According to the definition of the double-row monodromy matrix, we write
it explicitly as

\begin{eqnarray}
{\cal {T}}(\lambda )
&=&\left( \begin{array}{ccc}
{\cal {A}}_{11}(\lambda )&{\cal {A}}_{12}(\lambda ) &
{\cal {B}}_1(\lambda )\\
{\cal {A}}_{21}(\lambda )&{\cal {A}}_{22}(\lambda )
&{\cal {B}}_2(\lambda )\\
{\cal {C}}_1(\lambda )&{\cal {C}}_2(\lambda )&{\cal {D}}
(\lambda )\end{array}
\right) 
\nonumber \\
&=&T(\lambda )K(\lambda )T^{-1}(-\lambda )
\nonumber \\&=&
\left( \begin{array}{ccc}
A_{11}(\lambda )&A_{12}(\lambda ) &B_1(\lambda )\\
A_{21}(\lambda )&A_{22}(\lambda )&B_2(\lambda )\\
C_1(\lambda )&C_2(\lambda )&D(\lambda )\end{array}
\right)\times 
\left( \begin{array}{ccc}
k_1(\lambda )&0&0\\
0&k_2(\lambda )&0\\
0&0&k_3(\lambda )\end{array}
\right)
\nonumber \\
&\times&
\left( \begin{array}{ccc}
\bar {A}_{11}(-\lambda )&\bar {A}_{12}(-\lambda ) &\bar {B}_1(-\lambda )\\
\bar {A}_{21}(-\lambda )&\bar {A}_{22}(-\lambda )&
\bar {B}_2(-\lambda )\\
\bar {C}_1(-\lambda )&\bar {C}_2(-\lambda )&\bar {D}(-\lambda )\end{array}
\right) .
\end{eqnarray}
For convenience, we can write the inverse of the row-to-row
monodromy matrix as
\begin{eqnarray}
T^{-1}_a(-\lambda )=L_{1a}(\lambda )L_{2a}(\lambda )\cdots L_{Na}(\lambda ),
\end{eqnarray}
where we have used the unitarity relation of the R-matrix and a whole
factor is omitted.

Define reference state in the $n$-th quantum space and the vacuum $|0>$ as:
\begin{eqnarray}
|0>_n=\left(\begin{array}{c}
0\\0\\1\end{array}\right), 
|0>=\otimes _{k=1}^N|0>_k.
\end{eqnarray}
By use of the definition of the
row-to-row monodromy matrix (19) and its inverse(63), we have
\begin{eqnarray}                                                 
&&A_{ab}(\lambda )|0>=\delta _{ab}sin^N(\lambda )|0>,
~~~D(\lambda )|0>=sin^N(\lambda +\eta )|0>,                    
\nonumber \\
&&B_a(\lambda )|0>=0,
~~~~C_a(\lambda )|0>\not= 0, \\
&&{\bar {A}}_{ab}(-\lambda )|0>=\delta _{ab}sin^N(\lambda )|0>,
~~~{\bar {D}}(-\lambda )|0>=sin^N(\lambda +\eta )|0>,
\nonumber \\
&&{\bar {B}}_a(-\lambda )|0>=0,
~~~~{\bar {C}}_a(\lambda )|0>\not= 0. 
\end{eqnarray}
So, with the help of ${\cal {T}}$'s definition relation (62),
we can show that
\begin{eqnarray}
{\cal {D}}(\lambda )|0>&=&k_3(\lambda )sin^{2N}(\lambda +\eta )|0>,
\nonumber \\
\tilde {\cal {A}}_{ab}(\lambda )|0>&=&0, ~~~a\not= b,       
\nonumber \\                                                
{\cal {B}}_a(\lambda )|0>&=&0,
\nonumber \\
{\cal {C}}_a(\lambda )|0>&\not= &0.
\end{eqnarray}
To obtain the actions of operator $\tilde {\cal {A}}_{aa}$ on the
vacuum state, we use the following relation
obtained from the Yang-Baxter relation
\begin{eqnarray}
&&[T^{-1}(-\lambda )]_{a_2}^{b_2}
R(2\lambda )_{a_1b_2}^{b_1c_2}T(\lambda )_{b_1}^{c_1}
(-1)^{(\epsilon _{b_1}+\epsilon _{c_1})\epsilon _{c_2}}
\nonumber \\
&=&T(\lambda )_{a_1}^{b_1}R(2\lambda )_{b_1a_2}^{c_1b_2}
[T^{-1}(-\lambda )]_{b_2}^{c_2}(-1)^{(\epsilon _{a_1}
+\epsilon _{b_1})\epsilon _{a_2}}.
\end{eqnarray}
Actually, we have already used it to obtain the results
$\tilde {\cal {A}}_{ab}(\lambda )|0>=0, ~~a\not= b$.

Then, we have
\begin{eqnarray}
\tilde {\cal {A}}_{11}(\lambda )|0>=
\left[ k_1(\lambda )-k_3(\lambda )\frac {sin(\eta )e^{-2i\lambda }}
{sin(2\lambda +\eta )}\right] sin^{2N}(\lambda )|0>
\equiv W_1(\lambda )sin^{2N}(\lambda )|0>.
\end{eqnarray}
For case I, II reflecting K-matrix, we have a same $W_1$ which takes
the form:
\begin{eqnarray}
{\rm For}~K_I~{\rm and}~K_{II}:~~
W_1(\lambda )=e^{-2i\lambda }\frac {sin(2\lambda )sin(\xi +\lambda +\eta )}
{sin(2\lambda +\eta )}.
\end{eqnarray}
Similarly, we have
\begin{eqnarray}
\tilde {\cal {A}}_{22}(\lambda )|0>=
\left[ k_2(\lambda )-k_3(\lambda )\frac {sin(\eta )e^{-2i\lambda }}
{sin(2\lambda +\eta )}\right] sin^{2N}(\lambda )|0>
\equiv W_2(\lambda )sin^{2N}(\lambda )|0>.
\end{eqnarray}
For case I, II reflecting K-matrix, $W_2$ take the following forms
respectively:
\begin{eqnarray}
{\rm For}~K_I:~~~
W_2(\lambda )&=&e^{-2i\lambda }\frac {sin(2\lambda )sin(\xi +\lambda +\eta )}
{sin(2\lambda +\eta )},
\\
{\rm For}~K_{II}:~~~
W_2(\lambda )&=&e^{i\eta }\frac {sin(2\lambda )sin(\xi -\lambda )}
{sin(2\lambda +\eta )}.
\end{eqnarray}

\subsection{Bethe ansatz}

We construct a set 
of the eigenvectors of the transfer matrix with reflecting boundary
conditions as 
\begin{eqnarray}
{\cal {C}}_{d_1}(\mu _1)
{\cal {C}}_{d_2}(\mu _2)\cdots {\cal {C}}_{d_n}(\mu _n)|0>F^{d_1\cdots d_n}.
\end{eqnarray}
Here $F^{d_1\cdots d_n}$ is a function of the spectral parameters $\mu _j$.
Acting the transfer matrix on this eigenvectors, we find the
eigenvalues $\Lambda (\lambda )$
of the transfer matrix $t(\lambda )$
and a set of Bethe ansatz equations.
This technique is standard for the algebraic Bethe ansatz method.
Act first ${\cal {D}}$ on the
eigenvector defined above, use next the commutation relation (59), 
consider the value of ${\cal {D}}$ acting on the vacuum state (67). 
Then we have
\begin{eqnarray}
&&{\cal {D}}(\lambda )
{\cal {C}}_{d_1}(\mu _1)
{\cal {C}}_{d_2}(\mu _2)\cdots {\cal {C}}_{d_n}(\mu _n)|0>F^{d_1\cdots d_n}
\nonumber \\
&=&k_3(\lambda )sin^{2L}(\lambda +\eta )
\prod _{i=1}^n
\frac {sin(\lambda +\mu _i)sin(\lambda -\mu _i-\eta )}
{sin(\lambda +\mu _i+\eta )sin(\lambda -\mu _i)}
{\cal {C}}_{d_1}(\mu _1)
{\cal {C}}_{d_2}(\mu _2)\cdots {\cal {C}}_{d_n}(\mu _n)|0>F^{d_1\cdots d_n}
+u.t.,
\end{eqnarray}
where $u.t.$ means the unwanted terms. 

We act $\tilde {\cal {A}}_{aa}(\lambda )$ on the assumed eigenvector (74).
Using repeatedly the commutation relations (60), we have
\begin{eqnarray}
&&\tilde {\cal {A}}_{aa}(\lambda )
{\cal {C}}_{d_1}(\mu _1)
{\cal {C}}_{d_2}(\mu _2)\cdots {\cal {C}}_{d_n}(\mu _n)|0>F^{d_1\cdots d_n}
\nonumber \\
&=&\prod _{i=1}^n
\frac {1}
{sin(\lambda -\mu _i)sin(\lambda +\mu _i+\eta )}
r_{12}(\lambda +\mu _1+\eta )_{ac_1}^{a_1e_1}
r_{21}(\lambda -\mu _1)_{b_1e_1}^{ad_1}
\nonumber \\
&&r_{12}(\lambda +\mu _2+\eta )_{a_1c_2}^{a_2e_2}
r_{21}(\lambda -\mu _2)_{b_2e_2}^{b_1d_2}
\cdots 
r_{12}(\lambda +\mu _n+\eta )_{a_{n-1}c_n}^{a_ne_n}
r_{21}(\lambda -\mu _n)_{b_ne_n}^{b_{n-1}d_n}
\nonumber \\
&&\times {\cal {C}}_{c_1}(\mu _1)\cdots 
{\cal {C}}_{c_n}(\mu _n)\tilde {\cal {A}}_{a_nb_n}(\lambda )|0>
F^{d_1\cdots d_n}
+u.t.. 
\end{eqnarray}
Summarizing relations (67,69,71), we obtain 
\begin{eqnarray}
{\cal {A}}_{a_nb_n}(\lambda )|0>=\delta _{a_nb_n}W_{a_n}(\lambda )
sin^{2L}(\lambda )|0>.
\end{eqnarray}
We can rewrite the transfer matrix as  
\begin{eqnarray}
t(\lambda )&=&-k_1^+(\lambda )\tilde {\cal {A}}_{11}(\lambda )
-k_2^+(\lambda )\tilde {\cal {A}}_{22}(\lambda )+U_3^+(\lambda ) 
{\cal {D}}(\lambda )
\nonumber \\
&=&-k_a^+(\lambda ) 
\tilde {\cal {A}}_{aa}(\lambda )
+U_3^+(\lambda ) 
{\cal {D}}(\lambda ).
\end{eqnarray}
Thus the eigenvalue of the transfer matrix with reflecting boundary
condition is written as
\begin{eqnarray}
&&t(\lambda )
{\cal {C}}_{d_1}(\mu _1)
{\cal {C}}_{d_2}(\mu _2)\cdots {\cal {C}}_{d_n}(\mu _n)|0>F^{d_1\cdots d_n}
\nonumber \\
&=&U_3^+(\lambda )k_3(\lambda )sin^{2N}(\lambda +\eta )
\prod _{i=1}^n
\frac {sin(\lambda +\mu _i)(sin\lambda -\mu _i-\eta )}
{sin(\lambda +\mu _i+\eta )sin(\lambda -\mu _i)}
{\cal {C}}_{d_1}(\mu _1)
\cdots {\cal {C}}_{d_n}(\mu _n)|0>F^{d_1\cdots d_n}
\nonumber \\
&&+sin^{2N}(\lambda )
\prod _{i=1}^n
\frac {1}
{sin(\lambda -\mu _i)sin(\lambda +\mu _i+\eta )}
{\cal {C}}_{c_1}(\mu _1)
\cdots {\cal {C}}_{c_n}(\mu _n)|0>
t^{(1)}(\lambda )^{c_1\cdots c_n}_{d_1\cdots d_n}
F^{d_1\cdots d_n}
\nonumber \\
&&+u.t.,
\end{eqnarray}
where $t^{(1)}(\lambda )$ is the so-called nested transfer matrix, and
with the help of the relation (76), it can be defined as
\begin{eqnarray}
t^{(1)}(\lambda )^{c_1\cdots c_n}_{d_1\cdots d_n}
&=&-k_a^+(\lambda )
\left\{ r(\lambda +\mu _1+\eta )_{ac_1}^{a_1e_1}
r(\lambda +\mu _2+\eta )_{a_1c_2}^{a_2e_2}\cdots
r(\lambda +\mu _1+\eta )_{a_{n-1}c_n}^{a_ne_n}\right\}
\nonumber \\
&&\delta _{a_nb_n}W_{a_n}(\lambda )
\left\{ r_{21}(\lambda -\mu _n)_{b_ne_n}^{b_{n-1}d_n}
\cdots
r_{21}(\lambda -\mu _2)_{b_2e_2}^{b_1d_2}
r_{21}(\lambda -\mu _1)_{b_1e_1}^{ad_1}\right\} .
\end{eqnarray}
We find that this nested transfer matrix can be defined as
a transfer matrix with reflecting boundary conditions corresponding
to the anisotropic case
\begin{eqnarray}
t^{(1)}(\lambda )=str{K^{(1)}}^+(\tilde {\lambda })
T^{(1)}(\tilde {\lambda }, \{ \tilde {\mu }_i\} )
K^{(1)}(\tilde {\lambda })
{T^{(1)}}^{-1}(-\tilde {\lambda }, 
\{ \tilde {\mu }_i\} )
\end{eqnarray}
with the grading $\epsilon _1=\epsilon _2=1$. Here, we denote
$\tilde {\lambda }=\lambda +{\eta \over 2}, \tilde {\xi }=\xi +
{\eta \over 2},
\tilde {\xi }^+=\xi ^+-{\eta \over 2}$, and the same notation will be
used, for instance, $\tilde {\mu }=\mu +{\eta \over 2}$.
Explicitly we have 
\begin{eqnarray}
{K^{(1)}}^+_I(\tilde {\lambda })
=sin(\tilde {\xi }^+-\tilde {\lambda }+\eta )
e^{i(2\tilde {\lambda }-\eta )}
\left( \begin{array}{cc}
e^{i\eta }&\\
&e^{-i\eta }\end{array}\right) 
\end{eqnarray}
and
\begin{eqnarray}
{K^{(1)}}^+_{II}(\tilde {\lambda })=\left(
\begin{array}{cc}
sin(\tilde {\xi }^+-\tilde {\lambda }-\eta )e^{2i\tilde {\lambda }}&\\
&sin(\tilde {\xi }^++\tilde {\lambda }-\eta )\end{array}
\right) 
\end{eqnarray}
corresponding to $K_I^+$ and $K_{II}^+$ respectively. 
We also have 
\begin{eqnarray}
K_{I}^{(1)}(\tilde {\lambda })
&=&e^{-i(2\tilde {\lambda }-\eta )}
\frac {sin(2\tilde {\lambda }-\eta )sin(\tilde {\xi }+\tilde {\lambda })}
{sin(2\tilde {\lambda })}\cdot id,
\\
K_{II}^{(1)}(\tilde {\lambda })
&=&\frac {sin(2\tilde {\lambda }-\eta )e^{i\eta }}
{sin(2\tilde {\lambda })}
\left( \begin{array}{cc}
sin(\tilde {\xi }+\tilde {\lambda })e^{-2i\tilde {\lambda }}&\\
&sin(\tilde {\xi }-\tilde {\lambda })
\end{array}\right),
\end{eqnarray}
corresponding to $K_I$ and $K_{II}$.
The row-to-row monodromy matrix 
$T^{(1)}(\tilde {\lambda }, \{ \tilde {\mu }_i\} )$ (corresponding
to the periodic boundary condition) is defined as
\begin{eqnarray}
T^{(1)}_{aa_n}(\tilde {\lambda }, 
\{ \tilde {\mu }_i\} )_{c_1\cdots c_n}^{e_1\cdots e_n}
&=&
r(\tilde {\lambda }+\tilde {\mu }_1)_{ac_1}^{a_1e_1}
r(\tilde {\lambda }+\tilde {\mu }_2)_{a_1c_2}^{a_2e_2}\cdots
r(\tilde {\lambda }+\tilde {\mu }_1)_{a_{n-1}c_n}^{a_ne_n}
\nonumber \\
&=&L_1^{(1)}(\tilde {\lambda }+\tilde {\mu }_1)
L_2^{(1)}(\tilde {\lambda }+\tilde {\mu }_2)\cdots
L_n^{(1)}(\tilde {\lambda }+\tilde {\mu }_1).
\end{eqnarray}
The L-operator takes the form
\begin{eqnarray}
L^{(1)}_k(\tilde {\lambda })=
\left(
\begin{array}{cc}
b(\lambda )-(b(\lambda )-a(\lambda ))e_k^{11}&
-c_-(\lambda )e_n^{21}\\
-c_+(\lambda )e_n^{12} &b(\lambda )-(b(\lambda )-a(\lambda ))e_n^{22}
\end{array}\right).                                         
\end{eqnarray}
And we also have
\begin{eqnarray}
{T^{(1)}}^{-1}(-\tilde {\lambda }, 
\{ \tilde {\mu }_i\} )
&=&
r_{21}(\tilde {\lambda }-\tilde {\mu }_n)_{b_ne_n}^{b_{n-1}d_n}
\cdots
r_{21}(\tilde {\lambda }-\tilde {\mu }_2)_{b_2e_2}^{b_1d_2}
r_{21}(\tilde {\lambda }-\tilde {\mu }_1)_{b_1e_1}^{ad_1}
\nonumber \\
&=&{L_n^{(1)}}^{-1}(-\tilde {\lambda }+\tilde {\mu }_n )
\cdots 
{L_2^{(1)}}^{-1}(-\tilde {\lambda }+\tilde {\mu }_2 )
{L_1^{(1)}}^{-1}(-\tilde {\lambda }+\tilde {\mu }_1 ),
\end{eqnarray}
where we have used the 
unitarity relation of the r-matrix 
$r_{12}(\lambda )r_{21}(-\lambda )=sin(\eta -\lambda)sin(\eta +\lambda )
\cdot id.$.

In this section, we show that a problem to find the eigenvalue 
of the original transfer matrix $t(\lambda )$
reduces to a problem to find the eigenvalue of the nested transfer
matrix $t^{(1)}(\lambda )$. In relation (79), one can see that
besides the wanted term which gives the eigenvalue, we also
have the unwanted terms which must be cancelled so that the assumed
eigenvector is indeed the eigenvector of the transfer matrix. With
the help of the symmetry property (58) of the assumed eigenvector (74),
we find that,
if $\mu _1, \cdots, \mu _n$ satisfy the following Bethe ansatz equations,
the unwanted terms will vanish.
\begin{eqnarray}
U_3^+(\mu _j)k_3(\mu _j )sin^{2N}(\mu _j +\eta )
\prod _{i=1,\not= j}^n
sin(\mu _j +\mu _i)sin(\mu _j-\mu _i-\eta )
=-{sin^{2N}(\mu _j)}
\Lambda ^{(1)}({\mu }_j)
\nonumber \\
j=1, 2,\cdots ,n.
\end{eqnarray}
Here we have used the notation $\Lambda ^{(1)}(\lambda )$
to denote the eigenvalue
of the nested transfer matrix $t^{(1)}(\lambda )$.

Thus what we should do next is to find the eigenvalue of the nested
transfer matrix $t^{(1)}$.

\subsection{The nested algebraic Bethe ansatz method}
We expect that the eigenvalue of the nested transfer matrix can be solved
similarly as that of the original transfer matrix.
So, we should first prove that the above defined nested transfer matrix
indeed constitutes a commuting family.
Note that the grading is $\epsilon _1=\epsilon _2=1$. Actually, because
all grading is Fermionic, the graded method is simply the same
as the usual one.

We note that the r-matrix satisfies the unitarity and
and cross-unitarity relations:
\begin{eqnarray}
&&r_{12}(\lambda )r_{21}(-\lambda )=sin(\eta +\lambda )sin(\eta -\lambda )
\cdot id.,
\\
&&r_{12}^{st_1}(2\eta -\lambda )M_1^{(1)}r_{21}^{st_1}(\lambda )
{M_1^{(1)}}^{-1}=sin(\lambda )sin(2\lambda -\eta )\cdot id..
\end{eqnarray}
The matrix $M^{(1)}$ is a diagonal matrix, $M^{(1)}=diag.(e^{2i\eta }, 1)$.

In order to prove the commutativity of the nested transfer matrices,
we need the reflection equation and the dual reflection equation,
which take the following forms
\begin{eqnarray}
r_{12}(\lambda -\mu )K_1^{(1)}(\lambda )r_{21}(\lambda +\mu )
K_2^{(1)}(\mu )
=K_2^{(1)}(\mu )r_{12}(\lambda +\mu )K_1^{(1)}(\lambda )
r_{21}(\lambda -\mu ),
\end{eqnarray}
\begin{eqnarray}
&&r_{12}(\mu -\lambda ){K_1^{(1)}}^+(\lambda )
M_1^{-1}r_{21}(2\eta -\lambda -\mu )
{K_2^{(1)}}^+(\mu )M_2^{-1}                                        
\nonumber \\
&=&{K_2^{(1)}}^+(\mu )M_2^{-1}r_{12}(2\eta -\lambda -\mu )
{K_1^{(1)}}^+(\lambda )M_1^{-1}r_{21}(\mu -\lambda ).
\end{eqnarray}

By a direct calculation, we can prove that the above defined
reflecting matrices $K_I^{(1)}$ and $K_{II}^{(1)}$
satisfy the reflection equation, and also ${K_I^{(1)}}^+$,
${K_{II}^{(1)}}^+$ satisfy the dual reflection equation.

We know that the following graded Yang-Baxter relation is satisfied:
\begin{eqnarray}
r(\lambda -\mu )L_1^{(1)}(\lambda )L_2^{(1)}(\mu )
=L_2^{(1)}(\mu )L_1^{(1)}(\lambda )r(\lambda -\mu )
\end{eqnarray}
Therefore, we also have the Yang-Baxter
relation for the row-to-row monodromy matrix 
\begin{eqnarray}
r(\lambda -\mu )
T^{(1)}_1(\lambda , \{ \mu _i\} )
T^{(1)}_2(\mu , \{ \mu _i\} )
=T^{(1)}_2(\mu , \{ \mu _i\} )
T^{(1)}_1(\lambda , \{ \mu _i\} )
r(\lambda -\mu ).
\end{eqnarray}
Since we alreay know $K^{(1)}$ satisfy the reflection
equation (92), we can show that the nested double-row monodromy matrix 
\begin{eqnarray}
{\cal {T}}^{(1)}(\lambda ,\{ \mu _i\} )\equiv 
T^{(1)}(\lambda ,\{ \mu _i\} )K^{(1)}(\lambda )
{T^{(1)}}^{-1}(-\lambda ,\{ \mu _i\} ) 
\end{eqnarray}
also satisfy the the reflection equation
\begin{eqnarray}                                               
&&r_{12}(\lambda -\mu ){\cal {T}}^{(1)}_1(\lambda ,\{ \mu _i\} )
r_{21}(\lambda +\mu )
{\cal {T}}^{(1)}_2(\mu ,\{ \mu _i\} )
\nonumber \\
&=&{\cal {T}}^{(1)}_2(\nu ,\{ \mu _i\} )
r_{12}(\lambda +\mu )
{\cal {T}}^{(1)}_1(\lambda ,\{ \mu _i\} )r_{21}(\lambda -\mu ).
\end{eqnarray}
Parallel to the procedures presented above, with the help of unitarity,
cross-unitarity relations, and
reflection equation, dual reflection equation, one can prove
the defined nested transfer matrix indeed constitutes a commuting family.

Now, let us use again the algebraic Bethe ansatz method to obtain
the eigenvalue $\Lambda ^{(1)}(\lambda )$ of the nested transfer matrix
$t^{(1)}(\lambda )$. We write the nested double-row monodromy matrix as
\begin{eqnarray}
{\cal {T}}^{(1)}(\lambda ,\{ \mu _i\} )
&=&\left( \begin{array}{cc}
{\cal {A}}^{(1)}(\lambda ) &{\cal {B}}^{(1)}(\lambda ) \\
{\cal {C}}^{(1)}(\lambda ) &{\cal {D}}^{(1)}(\lambda ) \end{array}
\right)  
\nonumber \\
&=&T^{(1)}(\lambda ,\{ \mu _i\} )K^{(1)}(\lambda )
{T^{(1)}}^{-1}(-\lambda ,\{ \mu _i\} ) 
\nonumber \\
&=&\left( \begin{array}{cc}
A^{(1)}(\lambda ) &B^{(1)}(\lambda ) \\
C^{(1)}(\lambda ) &D^{(1)}(\lambda ) \end{array}
\right) \left( 
\begin{array}{cc}k^{(1)}_1(\lambda ) &\\
&k_2^{(1)}(\lambda )\end{array}\right)
\left( \begin{array}{cc}
{\bar {A}}^{(1)}(-\lambda ) &\bar {B}^{(1)}(-\lambda ) \\
\bar {C}^{(1)}(-\lambda ) &\bar {D}^{(1)}(-\lambda ) \end{array}
\right). \nonumber \\  
\end{eqnarray}
For convenience, we introduce again a transformation 
\begin{eqnarray}
{\cal {A}}^{(1)}(\lambda )=\tilde {\cal {A}}^{(1)}(\lambda )
-\frac {sin(\eta )e^{-2i\lambda }}{sin(2\lambda -\eta )}
{\cal {D}}^{(1)}(\lambda ).
\end{eqnarray}
Because the nested double-row monodromy matrix satisfies
the reflection equation (97), we can find the following commutation
relations:
\begin{eqnarray}
{\cal {D}}^{(1)}(\lambda )
{\cal {C}}^{(1)}(\mu )
&=&\frac {sin(\lambda -\mu +\eta )sin(\lambda +\mu )}
{sin(\lambda -\mu )sin(\lambda +\mu -\eta )}
{\cal {C}}^{(1)}(\mu )
{\cal {D}}^{(1)}(\lambda )
\nonumber \\
&-&\frac {sin(2\mu )sin(\eta )e^{i(\lambda -\mu )}}
{sin(\lambda -\mu )sin(2\mu -\eta )}
{\cal {C}}^{(1)}(\lambda )
{\cal {D}}^{(1)}(\mu )
+\frac {sin(\eta )e^{i(\lambda +\mu )}}{sin(\lambda +\mu -\eta )}
{\cal {C}}^{(1)}(\lambda )
\tilde {\cal {A}}^{(1)}(\mu ),
\end{eqnarray}

\begin{eqnarray}
\tilde {\cal {A}}^{(1)}(\lambda )
{\cal {C}}^{(1)}(\mu )
&=&\frac {sin(\lambda -\mu -\eta )sin(\lambda +\mu -2\eta )}
{sin(\lambda -\mu )sin(\lambda +\mu -\eta )}
{\cal {C}}^{(1)}(\mu )
\tilde {\cal {A}}^{(1)}(\lambda )
\nonumber \\
&+&\frac {sin(\eta )sin(2\lambda -2\eta )
e^{-i(\lambda -\mu )}}
{sin(\lambda -\mu )sin(2\lambda -\eta )}
{\cal {C}}^{(1)}(\lambda )
\tilde {\cal {A}}^{(1)}(\mu )
\nonumber \\
&&-\frac {sin(2\mu )sin(2\lambda -2\eta )sin(\eta )e^{-i(\lambda +\mu )}}
{sin(\lambda +\mu -\eta )sin(2\lambda -\eta )sin(2\mu -\eta )}
{\cal {C}}^{(1)}(\lambda )
{\cal {D}}^{(1)}(\mu ),
\end{eqnarray}

\begin{eqnarray}
{\cal {C}}^{(1)}(\lambda )
{\cal {C}}^{(1)}(\mu )
&=&
{\cal {C}}^{(1)}(\mu )
{\cal {C}}^{(1)}(\lambda  ).
\end{eqnarray}
As the reference states for the nesting, we choose
\begin{eqnarray}
|0>^{(1)}_k=\left( \begin{array}{c}0\\1\end{array}\right),
|0>^{(1)}=\otimes _{k=1}^n|0>_k^{(1)}.
\end{eqnarray}
With the help of the definition (86, 88), 
we know the actions of the nested monodromy matrix and the inverse
of the monodromy matrix on the reference state
\begin{eqnarray}
T^{(1)}(\lambda ,\{ \mu _i\} )|0>^{(1)}
&=&
\left( \begin{array}{cc}
A^{(1)}(\tilde {\lambda }) &B^{(1)}(\tilde {\lambda }) \\
C^{(1)}(\tilde {\lambda }) &D^{(1)}(\tilde {\lambda }) \end{array}
\right)|0>^{(1)}
\nonumber \\
&=& 
\left( \begin{array}{cc}
\prod _{i=1}^nsin(\tilde {\lambda }+\tilde {\mu }_i) &0 \\
C^{(1)}(\tilde {\lambda }) &
\prod _{i=1}^nsin(\tilde {\lambda }+\tilde {\mu }_i-\eta )
\end{array}
\right)|0>^{(1)}, 
\nonumber \\
\\
{T^{(1)}}^{-1}(-\lambda ,\{ \mu _i\} )|0>^{(1)} 
&=&\left( 
\begin{array}{cc}
\bar {A}^{(1)}(\tilde {\lambda }) &\bar {B}^{(1)}(\tilde {\lambda }) \\
\bar {C}^{(1)}(\tilde {\lambda }) &
\bar {D}^{(1)}(\tilde {\lambda }) \end{array}
\right)|0>^{(1)}
\nonumber \\
&=& 
\left(\begin{array}{cc}
\prod _{i=1}^nsin(\tilde {\lambda }-\tilde {\mu }_i) &0 \\
C^{(1)}(\tilde {\lambda }) &
\prod _{i=1}^nsin(\tilde {\lambda }-\tilde {\mu }_i-\eta )
\end{array}
\right)|0>^{(1)} .
\nonumber \\                                               
\end{eqnarray}                                             

Repeating almost the same calculation in the former sections,
we obtain the results of the nested double-row monodromy matrix acting
on the nested vacuum state $|0>^{(1)}$, 
\begin{eqnarray}
{\cal {B}}^{(1)}(\tilde {\lambda })|0>^{(1)}=0,~~~         
{\cal {C}}^{(1)}(\tilde {\lambda })|0>^{(1)}\not=0, 
\end{eqnarray}
\begin{eqnarray}
{\cal {D}}^{(1)}(\tilde {\lambda })|0>^{(1)}= 
U_2(\tilde {\lambda })\prod _{i=1}^n
[sin(\tilde {\lambda }+\tilde {\mu }_i-\eta )
sin(\tilde {\lambda }-\tilde {\mu }_i-\eta )]
|0>^{(1)}.
\end{eqnarray}
Here we use the notation $U_2=k^{(1)}_2$, 
\begin{eqnarray}
U_2(\tilde {\lambda })=
e^{-i(2\tilde {\lambda }-\eta )}
\frac {sin(2\tilde {\lambda }-\eta )sin(\tilde {\lambda }
+\tilde {\xi })}{sin(2\tilde {\lambda })}
\end{eqnarray}
for $K_I$ case.
\begin{eqnarray}
U_2(\tilde {\lambda })=\frac {e^{i\eta }
sin(2\tilde {\lambda }-\eta )
sin(\tilde {\xi }-\tilde {\lambda })}
{sin(2\tilde {\lambda })}
\end{eqnarray}
for $K_{II}$ case.

Using the Yang-Baxter relation, we also have 
\begin{eqnarray}
{\cal {A}}^{(1)}(\tilde {\lambda })|0>^{(1)}
&=&k_1^{(1)}(\tilde {\lambda })A^{(1)}(\tilde {\lambda })
\bar {A}^{(1)}(-\tilde {\lambda })|0>^{(1)}
\nonumber \\
&&+k_2^{(1)}(\tilde {\lambda })
\frac {b(2\tilde {\lambda })}{a(2\tilde {\lambda })-
b(2\tilde {\lambda })}
[A^{(1)}(\tilde {\lambda })
\bar {A}^{(1)}(-\tilde {\lambda })
-\bar {D}^{(1)}(-\tilde {\lambda })
D^{(1)}(\tilde {\lambda })]|0>^{(1)}
\nonumber \\
&=&[k_1^{(1)}(\tilde {\lambda })
+k_2^{(1)}(\tilde {\lambda })
\frac {sin(\eta )e^{-2i\lambda }}{sin(2\lambda -\eta )}
]
\prod _{i=1}^n[sin(\tilde {\lambda }+\tilde {\mu }_i)
sin(\tilde {\lambda }-\tilde {\mu }_i)]|0>^{(1)}
\nonumber \\
&&-\frac {sin(\eta )e^{-2i\lambda }}{sin(2\tilde {\lambda }-\eta )}
{\cal {D}}^{(1)}(\tilde {\lambda })|0>^{(1)} .
\end{eqnarray}
With the help of the transformation (99), we find
\begin{eqnarray}
\tilde {\cal {A}}^{(1)}(\tilde {\lambda })|0>^{(1)}
=U_1(\tilde {\lambda })
\prod _{i=1}^n[sin(\tilde {\lambda }+\tilde {\mu }_i)
sin(\tilde {\lambda }-\tilde {\mu }_i)]|0>^{(1)},
\end{eqnarray}
where we denote
\begin{eqnarray}
U_1(\tilde {\lambda })=k_1^{(1)}(\tilde {\lambda })
+k_2^{(1)}(\tilde {\lambda })
\frac {sin(2\eta e^{-2i\tilde {\lambda }})}{
sin(2\tilde {\lambda }-\eta )}.
\end{eqnarray}
Here $U_1$ takes the following form explicitly:
For $K_I(\lambda )$,
$U_1(\tilde {\lambda })=e^{-2i\tilde {\lambda }}
sin(\tilde {\lambda }+\tilde {\xi })$;
For $K_{II}(\lambda )$:
$U_1(\tilde {\lambda })=
e^{-i(2\tilde {\lambda }-\eta )}
sin(\tilde {\lambda }+\tilde {\xi }-\eta )$.

The nested transfer matrix takes the form
\begin{eqnarray}
t^{(1)}(\tilde {\lambda })&=&strK^{(1)}(\tilde {\lambda })
{\cal {T}}^{(1)}(\tilde {\lambda })
\nonumber \\
&=&-{k_1^{(1)}}^+(\tilde {\lambda }){\cal {A}}^{(1)}(\tilde {\lambda })
-{k_2^{(1)}}^+(\tilde {\lambda }){\cal {D}}^{(1)}(\tilde {\lambda })
\nonumber \\
&=&-U_1^+(\tilde {\lambda })
\tilde {\cal {A}}^{(1)}(\tilde {\lambda })
-U_2^+(\tilde {\lambda })
{\cal {D}}^{(1)}(\tilde {\lambda }),
\end{eqnarray}
where we denote $U_1^+={k_1^{(1)}}^+$, 
\begin{eqnarray}
U_2^+(\lambda )={k_2^{(1)}}^+(\lambda )-
\frac {sin(\eta )e^{-2i\tilde {\lambda }}}
{sin(2\lambda -\eta )}{k_1^{(1)}}^+(\lambda ),
\end{eqnarray}
that means:

For $K^+_I$ case:
\begin{eqnarray}
U_1^+(\tilde {\lambda })&=&sin(\tilde {\xi }^+-\tilde {\lambda }+\eta ),
\\
U_2^+(\tilde {\lambda })&=&
\frac {sin(2\tilde {\lambda }-2\eta )
sin(\tilde {\xi }^+-\tilde {\lambda }+\eta )
}{sin(2\tilde {\lambda }-\eta )}e^{i(2\tilde {\lambda }-\eta )}.
\end{eqnarray}
\indent For $K^+_{II}$ case:
\begin{eqnarray}
U_1^+(\tilde {\lambda })&=&sin(\tilde {\xi }^+-\tilde {\lambda }+\eta )
e^{2i\tilde {\lambda }},
\nonumber \\
U_2^+(\tilde {\lambda })&=&
\frac {sin(\tilde {\lambda }+\tilde {\xi }^+)
sin(2\tilde {\lambda }-2\eta )}
{sin(2\tilde {\lambda }-\eta )}.
\end{eqnarray}

Following the standard algebraic Bethe ansatz method, we assume that the
eigenvector of the nested transfer matrix is constructed as
${\cal {C}}(\tilde {\mu }^{(1)}_1)
{\cal {C}}(\tilde {\mu }^{(1)}_2)\cdots 
{\cal {C}}(\tilde {\mu }^{(1)}_m)|0>^{(1)}$. Acting the nested transfer
matrix on this eigenvector, using repeatedly the commutation
relations (100,101), we have the eigenvalue
\begin{eqnarray}
\Lambda ^{(1)}(\tilde {\lambda })
&=&-U_1^+(\tilde {\lambda })
U_1(\tilde {\lambda })
\prod _{i=1}^n[sin(\tilde {\lambda }+\tilde {\mu }_i)
sin(\tilde {\lambda }-\tilde {\mu }_i)]
\prod _{l=1}^m \left\{
\frac {sin(\tilde {\lambda }-\tilde {\mu }^{(1)}_l-\eta )
sin(\tilde {\lambda }+\tilde {\mu }^{(1)}_l-2\eta )}
{sin(\tilde {\lambda }-\tilde {\mu }^{(1)}_l)
sin(\tilde {\lambda }+\tilde {\mu }^{(1)}_l-\eta )}\right\} 
\nonumber\\ 
&&-U_2^+(\tilde {\lambda })
U_2(\tilde {\lambda })\prod _{i=1}^n
[sin(\tilde {\lambda }+\tilde {\mu }_i-\eta )
sin(\tilde {\lambda }-\tilde {\mu }_i-\eta )] 
\nonumber \\
&&\prod _{l=1}^m
\left\{ \frac {sin(\tilde {\lambda }-\tilde {\mu }^{(1)}_l+\eta )
sin(\tilde {\lambda }+\tilde {\mu }^{(1)}_l)}
{sin(\tilde {\lambda }-\tilde {\mu }^{(1)}_l)
sin(\tilde {\lambda }+\tilde {\mu }^{(1)}_l-\eta )}\right\} ,
\end{eqnarray}
where $\tilde {\mu }^{(1)}_1, \cdots ,\tilde {\mu }^{(1)}_m$
should satisfy the following Bethe ansatz equations
\begin{eqnarray}
&&\frac {U_1^+(\tilde {\mu }_j^{(1)})
U_1(\tilde {\mu }_j^{(1)})}
{U_2^+(\tilde {\mu }_j^{(1)})
U_2(\tilde {\mu }_j^{(1)})}
\frac {sin(2\tilde {\mu }_j^{(1)}-2\eta )}{sin(2\tilde {\mu }_j^{(1)})}
\prod _{i=1}^n
\frac {sin(\tilde {\mu }_j^{(1)}+\tilde {\mu }_i)
sin(\tilde {\mu }_j^{(1)}-\tilde {\mu }_i)}
{sin(\tilde {\mu }_j^{(1)}+\tilde {\mu }_i-\eta )
sin(\tilde {\mu }_j^{(1)}-\tilde {\mu }_i-\eta )} 
\nonumber \\
&=&
\prod _{l=1, \not=j}^m
\frac {sin(\tilde {\mu }_j^{(1)}-\tilde {\mu }^{(1)}_l+\eta )
sin(\tilde {\mu }_j^{(1)}+\tilde {\mu }^{(1)}_l)}
{sin(\tilde {\mu }_j^{(1)}-\tilde {\mu }^{(1)}_l-\eta )
sin(\tilde {\mu }_j^{(1)}+\tilde {\mu }^{(1)}_l-2\eta )},
~~~j=1, \cdots , m.
\end{eqnarray}
We already know the exact form of $\Lambda ^{(1)}$,
so we can change the former Bethe ansatz
equation presented in relation (89) as follows:
\begin{eqnarray}
1=\frac {U_2^+(\tilde {\mu }_j)U_2(\tilde {\mu }_j)}
{U_3^+(\tilde {\mu }_j)U_3(\tilde {\mu }_j)}
\frac {sin^{2N}(\mu _j)}{sin^{2N}(\mu _j+\eta )}
\prod _{l=1}^m
\frac {sin(\tilde {\mu }_j-\tilde {\mu }_l^{(1)}+\eta )
sin(\tilde {\mu }_j+\tilde {\mu }_l^{(1)})}
{sin(\tilde {\mu }_j-\tilde {\mu }_l^{(1)})
sin(\tilde {\mu }_j+\tilde {\mu }_l^{(1)}-\eta )},
\nonumber \\
j=1,\cdots ,n.
\end{eqnarray}
                                                  
The eigenvalue of the transfer matrix $t(\lambda )$
with reflecting boundary condition (46) is obtained as
\begin{eqnarray}
\Lambda (\lambda )
&=&U_3^+(\lambda )U_3(\lambda )sin^{2N}(\lambda +\eta )
\prod _{i=1}^n
\frac {sin(\lambda +\mu _i)sin(\lambda -\mu _i-\eta )}
{sin(\lambda +\mu _i+\eta )sin(\lambda -\mu _i)}
\nonumber \\
&+&sin^{2N}(\lambda )
\prod _{i=1}^n
\frac {1}
{sin(\lambda -\mu _i)sin(\lambda +\mu _i+\eta )}
\Lambda ^{(1)}(\tilde {\lambda }).
\end{eqnarray}
Here for convenience, we give a summary of the values $U$ and $U^+$. 

Case I:
\begin{eqnarray}
U_1^+(\tilde {\lambda })&=&sin(\tilde {\xi }^+-\tilde {\lambda }+\eta )
e^{2i\tilde {\lambda }},
\nonumber \\
U_2^+(\tilde {\lambda })&=&
\frac {sin(2\tilde {\lambda }-2\eta )
sin(\tilde {\xi }^+-\tilde {\lambda }+\eta )
}{sin(2\tilde {\lambda }-\eta )}e^{i(2\tilde {\lambda }-\eta )},
\nonumber \\
U_3^+(\lambda )&=&\frac {sin(2\lambda -\eta )
sin(\xi ^++\lambda +\eta )}{sin(2\lambda +\eta )}.
\end{eqnarray}

Case II:
\begin{eqnarray}
U_1^+(\tilde {\lambda })&=&sin(\tilde {\xi }^+-\tilde {\lambda }+\eta )
e^{2i\tilde {\lambda }},
\nonumber \\
U_2^+(\tilde {\lambda })&=&
\frac {sin(\tilde {\lambda }+\tilde {\xi }^+)sin(2\tilde {\lambda }-2\eta )}
{sin(2\tilde {\lambda }-\eta )},
\nonumber \\
U_3^+(\lambda )&=&\frac {sin(2\lambda -\eta )
sin(\xi ^++\lambda )}{sin(2\lambda +\eta }e^{i\eta }.
\end{eqnarray}

Case I:
\begin{eqnarray}
U_1(\tilde {\lambda })&=&sin(\tilde {\lambda }+\tilde {\xi })
e^{-2i\tilde {\lambda }},
\nonumber \\
U_2(\tilde {\lambda })&=&
e^{-i(2\tilde {\lambda }-\eta )}
\frac {sin(2\tilde {\lambda }-\eta )sin(\tilde {\lambda }
+\tilde {\xi })}{sin(2\tilde {\lambda })},
\nonumber \\
U_3(\lambda )&=&sin(\xi -\lambda ).
\end{eqnarray}

Case II:
\begin{eqnarray}
U_1(\tilde {\lambda })&=&sin(\tilde {\lambda }+\tilde {\xi }-\eta )
e^{-i(2\tilde {\lambda }-\eta )}.
\nonumber \\
U_2(\tilde {\lambda })&=&e^{2\eta }
\frac {sin(2\tilde {\lambda }-\eta )
sin(\tilde {\xi }-\tilde {\lambda })}{sin(2\tilde {\lambda })},
\nonumber \\
U_3(\lambda )&=&sin(\xi -\lambda ).
\end{eqnarray}
Since U and $U^+$ are independent of each other, there
are four combinations for $\{ U, U^+\} $ such as $\{I, I\}$,
$\{I, II\}$, $\{II, I\}$ and $\{II, II\}$.

In a special limit $\xi \rightarrow -i\infty $, the solution of the
reflection equation becomes identity, our result should be reduced
to the results obtained by Foerster and Karowski\cite{FK1}. And in
the rational limit, the results are equivalent to the previous
results\cite{FHS,AS}. 

\section{Algebraic Bethe ansatz for BFF grading}
\subsection{The first-level Bethe ansatz} 
For the case of BFF grading, the calculations proceed parallel to the case
of FFB. However, for the nested algebraic Bethe ansatz method,
the low-level r-matrix is BF grading which is significantly different
from the FF grading r-matrix. Actually, as we observed in the last section,
the graded method is the same as the usual method for FF grading r-matrix.
We shall study the supersymmetric $t-J$ model in the BFF grading.

The R-matrix is now
\begin{eqnarray}
R(\lambda )=\left(
\begin{array}{ccccccccc}
w(\lambda )&0&0&0&0&0&0&0&0\\                                        
0&b(\lambda )&0&c_-(\lambda )&0&0&0&0&0\\
0&0&b(\lambda )&0&0&0&c_-(\lambda )&0&0\\
0&c_+(\lambda )&0&b(\lambda )&0&0&0&0&0\\
0&0&0&0&a(\lambda )&0&0&0&0\\
0&0&0&0&0&b(\lambda )&0&-c_-(\lambda )&0\\
0&0&c_+(\lambda )&0&0&0&b(\lambda )&0&0\\
0&0&0&0&0&-c_+(\lambda )&0&b(\lambda )&0\\
0&0&0&0&0&0&0&0&a(\lambda )
\end{array}\right).
\end{eqnarray}
The diagonal solutions of the dual reflection equation are
\begin{eqnarray}
K^+_{I}(\lambda )&=&\left(
\begin{array}{ccc}
sin(\xi ^+-\lambda )e^{i(2\lambda -\eta )}&&\\
&sin(\xi ^+-\lambda )e^{i(2\lambda -\eta)}&\\
&&sin(\xi ^++\lambda -\eta )e^{-2i\eta }\end{array}\right),                    
\nonumber \\
K^+_{II}(\lambda )&=&\left(
\begin{array}{ccc}
sin(\xi ^+-\lambda )e^{i(2\lambda -\eta )}&&\\
&sin(\xi ^++\lambda -\eta )&\\
&&sin(\xi ^++\lambda -\eta )e^{-2i\eta }\end{array}\right),
\end{eqnarray}
where $\xi ^+$ is an arbitrary boundary parameter.

We still denote solution of the dual reflection equation $K^+$ and the
double-row monodromy matrix ${\cal {T}}$ respectively
in the following forms:
\begin{eqnarray}
K^+(\lambda )&=&diag.(k_1^+(\lambda ), k_2^+(\lambda ),
k_3^+(\lambda )),
\\
{\cal {T}}(\lambda )
&=&\left( \begin{array}{ccc}
{\cal {A}}_{11}(\lambda )&{\cal {A}}_{12}(\lambda ) &
{\cal {B}}_1(\lambda )\\
{\cal {A}}_{21}(\lambda )&{\cal {A}}_{22}(\lambda )
&{\cal {B}}_2(\lambda )\\
{\cal {C}}_1(\lambda )&{\cal {C}}_2(\lambda )&{\cal {D}}
(\lambda )\end{array}                                          
\right) .
\end{eqnarray}

In order to obtain the commutation relations, we need the following
transformation:
\begin{eqnarray}
{\cal {A}}(\lambda )_{ab}=\tilde {\cal {A}}(\lambda )_{ab}
-\delta _{ab}
\frac {e^{-2i\lambda }sin(\eta )}{sin(2\lambda -\eta )}
{\cal {D}}(\lambda ).
\end{eqnarray}
Because the double-row monodromy matrix satisfies the reflection equation,
we obtain the following
commutation relations after some tedious calculations:
\begin{eqnarray}
{\cal {C}}_{d_1}(\lambda ){\cal {C}}_{d_2}(\mu )
&=&\frac {r_{12}(\lambda -\mu )_{c_2c_1}^{d_2d_1}}{sin(\lambda -\mu -\eta )}
(-1)^{1+\epsilon _{d_1}+\epsilon _{c_2}+
\epsilon _{c_1}\epsilon _{c_2}}
{\cal {C}}_{c_2}(\mu ){\cal {C}}_{c_1}(\lambda ),
\end{eqnarray}

\begin{eqnarray}
{\cal {D}}(\lambda ){\cal {C}}_{d}(\mu )
&=&\frac {sin(\lambda +\mu )sin(\lambda -\mu +\eta )}
{sin(\lambda +\mu -\eta )sin(\lambda -\mu )}
{\cal {C}}_{d}(\mu )
{\cal {D}}(\lambda )
\nonumber \\
&&-\frac {sin(2\mu )sin(\eta )e^{i(\lambda -\mu }}
{sin(\lambda -\mu )sin(2\mu -\eta )}  
{\cal {C}}_{d}(\lambda )
{\cal {D}}(\mu )
+\frac {sin(\eta )e^{i(\lambda +\mu )}}
{sin(\lambda +\mu -\eta )}
{\cal {C}}_b(\lambda )\tilde {\cal {A}}_{bd}(\mu ),
\end{eqnarray}

\begin{eqnarray}
\tilde {\cal {A}}_{a_1d_1}(\lambda ){\cal {C}}_{d_2}(\mu )
&=&(-1)^{\epsilon _{a_1}+\epsilon _{d_1}+\epsilon _{c_1}\epsilon _{b_2}
+\epsilon _{d_1}\epsilon _{d_2}}
\frac {r_{12}(\lambda +\mu -\eta )_{a_1c_2}^{c_1b_2}
r_{21}(\lambda -\mu )_{b_1b_2}^{d_1d_2}}
{sin(\lambda +\mu -\eta )sin(\lambda -\mu )}                
{\cal {C}}_{c_2}(\mu )
\tilde {\cal {A}}_{c_1b_1}(\lambda )
\nonumber \\
&&-(-1)^{\epsilon _{a_1}(1+\epsilon _{b_1})+\epsilon _{d_1}}
\frac {sin(\eta )e^{-i(\lambda -\mu )}}
{sin(\lambda -\mu )sin(2\lambda -\eta )}  
r_{12}(2\lambda -\eta )_{a_1b_1}^{b_2d_1}
{\cal {C}}_{b_1}(\lambda )
\tilde {\cal {A}}_{b_2d_2}(\mu )
\nonumber \\
&&+(-1)^{\epsilon _{d_1}+\epsilon _{a_1}(\epsilon _{d_1}+\epsilon _{d_2})}
\frac {sin(2\mu )sin(\eta )e^{-i(\lambda +\mu )}}
{sin(\lambda +\mu -\eta )sin(2\lambda -\eta )sin(2\mu -\eta )}
\nonumber \\
&&~~~~~\times r_{12}(2\lambda +\eta )_{a_1b_2}^{d_2d_1}
{\cal {C}}_{b_2}(\lambda ){\cal {D}}(\mu ).
\end{eqnarray}
Here the indices take values 1,2, and the Grassmann parities
are BF, $\epsilon _1=0, \epsilon _2=1$. The r-matrix is defined as
\begin{eqnarray}
r_{12}(\lambda )=\left( \begin{array}{cccc}
sin(\lambda +\eta )&0&0&0\\
0&sin(\lambda )&sin(\eta )e^{-i\lambda }&0\\
0&sin(\eta )e^{i\lambda }&sin(\lambda )&0\\
0&0&0&sin(\lambda -\eta )\end{array}\right). 
\end{eqnarray}
The elements of the r-matrix are equal to
those of the original R-matrix when its indices just take values 1,2,
and the Grassmann parities also remain the same as before if we just
take values 1,2. This r-matrix has the $su(1|1)$ symmetry.

Let the elements of the double-row monodromy matrix act on
the vacuum state $|0>$:
\begin{eqnarray}
{\cal {D}}(\lambda )|0>&=&U_3(\lambda )sin^{2N}(\lambda -\eta )|0>,
\nonumber \\
{\cal {A}}_{aa}(\lambda )|0>
&=&W_a(\lambda )sin^{2N}(\lambda )|0>,
\nonumber \\                                                 
\tilde {\cal {A}}_{ab}(\lambda )|0>&=&0, ~~~a\not= b       
\nonumber \\                                                
{\cal {B}}_a(\lambda )|0>&=&0,
\nonumber \\
{\cal {C}}_a(\lambda )|0>&\not= &0.
\end{eqnarray}
Here we have defined                                   
\begin{eqnarray}
U_3(\lambda )=k_3(\lambda ),
~~~~~                                          
W_a(\lambda )=k_a(\lambda )+\frac {sin(\eta )e^{-2i\eta }}
{sin(2\lambda -\eta )}k_3(\lambda ).
\end{eqnarray}
Substituting the exact forms of the reflecting type I and type II
K-matrices into the above relation, we have
\begin{eqnarray}
W_1^I(\lambda )&=&W_2^I(\lambda )=
\frac {sin(\lambda )
sin(\xi +\lambda -\eta )e^{-i2\lambda }}
{sin(2\lambda -\eta )},
\nonumber \\
W_1^{II}(\lambda )&=&
\frac {sin(2\lambda )sin(\xi +\lambda -\eta )
e^{-i2\lambda }}
{sin(2\lambda -\eta )},
\nonumber \\
W_2^{II}(\lambda )&=&
\frac {sin(2\lambda )sin(\xi -\lambda )
e^{-i\eta }}
{sin(2\lambda -\eta )}.
\end{eqnarray}

The transfer matrix with boundaries for BFF grading is written as
\begin{eqnarray}
t(\lambda )&=&k_1^+(\lambda ){\cal {A}}_{11}(\lambda )
-k_2^+(\lambda ){\cal {A}}_{22}(\lambda )                        
-k_3^+(\lambda ){\cal {D}}(\lambda )
\nonumber \\
&=&(-1)^{\epsilon _a}k_a^+(\lambda )\tilde {\cal {A}}_{aa}(\lambda )
+U_3^+(\lambda )
{\cal {D}}(\lambda ),
\end{eqnarray}
where $U_3^+$ is defined by 
\begin{eqnarray}
U_3^+(\lambda )\equiv
k_3^+(\lambda )+
\frac {e^{-2i\lambda }sin(\eta )}{sin(2\lambda -\eta )}
(k_1^+(\lambda )-k_2^+(\lambda )).
\end{eqnarray}
For type I, II solutions of the dual reflection equations $K^+$, we have
\begin{eqnarray}
U_3^+(\lambda )&=&k_3(\lambda )
=sin(\xi ^++\lambda -\eta )e^{-2i\eta }, ~~~{\rm for}~~K_I^+,
\\
U_3^+(\lambda )&=&sin(\xi ^++\lambda -2\eta )e^{-i\eta },
~~~~~~~{\rm for}~~K_{II}^+.
\end{eqnarray}
Using the standard algebraic Bethe ansatz method, acting the above
defined transfer matrix on the ansatz of eigenvector
${\cal {C}}_{d_1}(\mu _1)
{\cal {C}}_{d_2}(\mu _2)\cdots {\cal {C}}_{d_n}(\mu _n)|0>
F^{d_1\cdots d_n}$, we have
\begin{eqnarray}
&&t(\lambda )
{\cal {C}}_{d_1}(\mu _1)
{\cal {C}}_{d_2}(\mu _2)\cdots {\cal {C}}_{d_n}(\mu _n)|0>F^{d_1\cdots d_n}
\nonumber \\
&=&U_3^+(\lambda )U_3(\lambda )sin^{2N}(\lambda -\eta )
\prod _{i=1}^n
\frac {sin(\lambda +\mu _i)(sin\lambda -\mu _i+\eta )}
{sin(\lambda +\mu _i-\eta )sin(\lambda -\mu _i)}
{\cal {C}}_{d_1}(\mu _1)
\cdots {\cal {C}}_{d_n}(\mu _n)|0>F^{d_1\cdots d_n}
\nonumber \\
&+&sin^{2N}(\lambda )                                        
\prod _{i=1}^n
\frac {1}
{sin(\lambda -\mu _i)sin(\lambda +\mu _i-\eta )}
{\cal {C}}_{c_1}(\mu _1)
\cdots {\cal {C}}_{c_n}(\mu _n)|0>
t^{(1)}(\lambda )^{c_1\cdots c_n}_{d_1\cdots d_n}            
F^{d_1\cdots d_n}
\nonumber \\
&+&u.t.,
\end{eqnarray}
where the nested transfer matrix
$t^{(1)}(\lambda )$ is defined as
\begin{eqnarray}
t^{(1)}(\lambda )^{c_1\cdots c_n}_{d_1\cdots d_n}
&=&(-)^{\epsilon _{a}}k_a^+(\lambda )
\left\{ r(\lambda +\mu _1-\eta )_{ac_1}^{a_1e_1}
r(\lambda +\mu _2-\eta )_{a_1c_2}^{a_2e_2}\cdots
r(\lambda +\mu _1-\eta )_{a_{n-1}c_n}^{a_ne_n}\right\}
\nonumber \\
&&\delta _{a_nb_n}W_{a_n}(\lambda )
\left\{ r_{21}(\lambda -\mu _n)_{b_ne_n}^{b_{n-1}d_n}
\cdots
r_{21}(\lambda -\mu _2)_{b_2e_2}^{b_1d_2}
r_{21}(\lambda -\mu _1)_{b_1e_1}^{ad_1}
\right\} 
\nonumber \\
&&\times 
(-1)^{\sum _{i=1}^n(\epsilon _{a_i}+\epsilon _{b_i})(1+\epsilon _{e_i})}.
\end{eqnarray}
Here we have used
$\epsilon _a\epsilon _b=\epsilon _c\epsilon _d$
for a non-zero elements of the r-matrix
$r_{ab}^{cd}$.
We also know that for non-zero $r_{ab}^{cd}$, we have
$\epsilon _a+\epsilon _c=\epsilon _b+\epsilon _d$. Considering
$\epsilon _a+\epsilon _a=0$, we can write
\begin{eqnarray}                    
\epsilon _{a_i}+\epsilon _{b_i}&=&
\epsilon _{a_i}+2\epsilon _{a_{i+1}}+\cdots +2\epsilon _{a_{n-1}}+
2\epsilon _{b_{n-1}}
+\cdots
+2\epsilon _{b_{i+1}}+\epsilon _{b_i}
\nonumber \\
&=&\sum _{j=1}^{n-i}(\epsilon _{c_{i+j}}
+\epsilon _{d_{i+j}}),~~~~i=1,\cdots ,n-1,
\end{eqnarray}
which
\begin{eqnarray}
\sum _{i=1}^n(\epsilon _{a_i}+\epsilon _{b_i})(1+\epsilon _{e_i})
=\sum _{j=2}^n(\epsilon _{c_j}+\epsilon _{e_j})
\sum _{i=1}^{j-1}(1+\epsilon _{e_i})
+\sum _{j=2}^n(\epsilon _{d_j}+\epsilon _{e_j})
\sum _{i=1}^{j-1}(1+\epsilon _{e_i}).
\end{eqnarray}
Thus this nested transfer matrix can still be interpreted as
a transfer matrix with reflecting boundary conditions corresponding
to the anisotropic case
\begin{eqnarray}
t^{(1)}(\lambda )=str{K^{(1)}}^+(\tilde {\lambda })
T^{(1)}(\tilde {\lambda }, \{ \tilde {\mu }_i\} )
K^{(1)}(\tilde {\lambda })
{T^{(1)}}^{-1}(-\tilde {\lambda }, 
\{ \tilde {\mu }_i\} )
\end{eqnarray}                             
with the grading BF $\epsilon _1=0, \epsilon _2=1$, where we denote
$\tilde {x}=x-{\eta \over 2}, x=\lambda ,\mu ,\xi ,\xi ^+$.
According to the definition, we have nested reflecting matrices:
\begin{eqnarray}
K^{(1)}(\tilde {\lambda })&\equiv &\left( \begin{array}{cc}
W_1(\tilde {\lambda } +{1\over 2}\eta )&\\
&W_2(\tilde {\lambda }+{1\over 2}\eta )\end{array}\right)
\nonumber \\
&=&\left\{
\begin{array}{ll}
\frac {sin(2\tilde {\lambda }+\eta )sin(\tilde {\xi }+\tilde {\lambda })}
{sin(2\tilde {\lambda })}e^{-i(2\tilde {\lambda }+\eta )}
\cdot id. ,&{\rm For~Case~I},
\\
\frac {sin(2\tilde {\lambda }+\eta )e^{-i(2\tilde {\lambda }+\eta )}}
{sin(2\tilde {\lambda })}\cdot
diag.\left( sin(\tilde {\xi }+\tilde {\lambda }),
sin(\tilde {\xi }-\tilde {\lambda })e^{i2\tilde {\lambda }}\right),
&{\rm For~Case~II},
\end{array}
\right.
\end{eqnarray}
and
\begin{eqnarray}
{K^{(1)}}^+(\tilde {\lambda })&\equiv &\left( \begin{array}{cc}
k_1^+(\tilde {\lambda } +{1\over 2}\eta )&\\
&k_2^+(\tilde {\lambda }+{1\over 2}\eta )\end{array}\right)
\nonumber \\
&=&\left\{
\begin{array}{ll}
sin(\tilde {\xi }^+-\tilde {\lambda })
e^{i2\tilde {\lambda }}
\cdot id. ,&{\rm For~Case~I},
\\
diag.\left( sin(\tilde {\xi }-\tilde {\lambda })e^{i2\tilde {\lambda }},
sin(\tilde {\xi }+\tilde {\lambda })\right),
&{\rm For~Case~II}.
\end{array}
\right.
\end{eqnarray}
The row-to-row monodromy matrix
$T^{(1)}(\tilde {\lambda }, \{ \tilde {\mu }_i\} )$ and
${T^{(1)}}^{-1}(-\tilde {\lambda }, 
\{ \tilde {\mu }_i\} )$ are defined respectively as
\begin{eqnarray}                       
&&T^{(1)}_{aa_n}(\tilde {\lambda }, 
\{ \tilde {\mu }_i\} )_{c_1\cdots c_n}^{e_1\cdots e_n}
=
r(\tilde {\lambda }+\tilde {\mu }_1)_{ac_1}^{a_1e_1}
r(\tilde {\lambda }+\tilde {\mu }_2)_{a_1c_2}^{a_2e_2}\cdots
r(\tilde {\lambda }+\tilde {\mu }_1)_{a_{n-1}c_n}^{a_ne_n}
(-1)^{\sum _{j=2}^n(\epsilon _{c_i}+\epsilon _{e_i})
\sum _{i=1}^{j-1}(1+\epsilon _{e_i})}
\nonumber \\
&&=L^{(1)}(\tilde {\lambda }+\tilde {\mu }_1)_{ac_1}^{a_1e_1}
L^{(1)}(\tilde {\lambda }+\tilde {\mu }_2)_{a_1c_2}^{a_2e_2}
\cdots
L^{(1)}(\tilde {\lambda }+\tilde {\mu }_1)_{a_{n-1}c_n}^{a_ne_n}
(-1)^{\sum _{j=2}^n(\epsilon _{c_i}+\epsilon _{e_i})
\sum _{i=1}^{j-1}(1+\epsilon _{e_i})},
\end{eqnarray}
\begin{eqnarray}
{T^{(1)}}^{-1}(-\tilde {\lambda }, 
\{ \tilde {\mu }_i\} )
&=&r_{21}(\tilde {\lambda }-\tilde {\mu }_n)_{b_ne_n}^{b_{n-1}d_n}
\cdots
r_{21}(\tilde {\lambda }-\tilde {\mu }_2)_{b_2e_2}^{b_1d_2}
\nonumber \\
&&~~~~r_{21}(\tilde {\lambda }-\tilde {\mu }_1)_{b_1e_1}^{ad_1}
(-1)^{\sum _{j=2}^n(\epsilon _{d_i}+\epsilon _{e_i})
\sum _{i=1}^{j-1}(1+\epsilon _{e_i})}
\nonumber \\
&=&{L_n^{(1)}}^{-1}(-\tilde {\lambda }+\tilde {\mu }_n )
_{b_ne_n}^{b_{n-1}d_n}
\cdots 
{L_2^{(1)}}^{-1}(-\tilde {\lambda }+\tilde {\mu }_2 )_{b_2e_2}^{b_1d_2}
\nonumber \\
&&~~~{L_1^{(1)}}^{-1}(-\tilde {\lambda }+\tilde {\mu }_1 )_{b_1e_1}^{ad_1}
(-1)^{\sum _{j=2}^n(\epsilon _{d_i}+\epsilon _{e_i})
\sum _{i=1}^{j-1}(1+\epsilon _{e_i})},
\end{eqnarray}
where we have used the unitarity relation of the r-matrix 
$r_{12}(\lambda )r_{21}(-\lambda )=sin(\eta -\lambda)sin(\eta +\lambda )
\cdot id.$. The L-operator is obtained from the r-matrix and takes the form
\begin{eqnarray}
L^{(1)}_k(\tilde {\lambda })=
\left(
\begin{array}{cc}
b(\lambda )-(b(\lambda )-w(\lambda ))e_k^{11}&
c_-(\lambda )e_n^{21}\\
c_+(\lambda )e_n^{12} &b(\lambda )-(b(\lambda )-a(\lambda ))e_n^{22}
\end{array}\right).                                         
\end{eqnarray}
We find that the super tensor product in the above defined
monodromy matrix is different from the original definition.
Nevertheless, as in the periodic boundary condition case, we can
define another graded tensor product as follows \cite{EK}:
\begin{eqnarray}
{F\bar {\otimes }G}_{ac}^{bd}=F_a^bG_c^d(-1)^{(\epsilon _a+\epsilon _b)
(1+\epsilon _c)}.
\end{eqnarray}
Effectively the graded tensor product switches even and odd Grassmann
parities. The graded tensor product in the above monodromy matrices
follows the new defined rule.

The L-operator satisfies the following Yang-Baxter relation
\begin{eqnarray}
r(\lambda -\mu )_{a_1a_2}^{b_1b_2}L^{(1)}(\lambda )_{b_1}^{c_1}
L^{(1)}(\mu )_{b_2}^{c_2}(-1)^{(\epsilon _{b_1}+\epsilon _{c_1})
\epsilon _{b_2}}
=L^{(1)}(\mu )_{a_2}^{b_2}L^{(1)}(\lambda )_{a_1}^{b_1}
r(\lambda -\mu )_{b_1b_2}^{c_1c_2}(-1)^{(\epsilon _{a_1}+\epsilon _{b_1})
\epsilon _{b_2}}.
\end{eqnarray}
Multiplying both sides of this Yang-Baxter relation by
$(-1)^{(\epsilon _{a_1}+\epsilon _{c_1})}$, we obtain
\begin{eqnarray}
&&\hat {r}(\lambda -\mu )_{a_1a_2}^{b_1b_2}L^{(1)}(\lambda )_{b_1}^{c_1}
L^{(1)}(\mu )_{b_2}^{c_2}(-1)^{(\epsilon _{b_1}+\epsilon _{c_1})
(1+\epsilon _{b_2})}
\nonumber \\
&=&L^{(1)}(\mu )_{a_2}^{b_2}L^{(1)}(\lambda )_{a_1}^{b_1}
\hat {r}(\lambda -\mu )_{b_1b_2}^{c_1c_2}
(-1)^{(\epsilon _{a_1}+\epsilon _{b_1})(1+\epsilon _{b_2})}.
\end{eqnarray}
This is just the graded Yang-Baxter relation in the newly defined
graded tensor product. And we have another r-matrix
\begin{eqnarray}
\hat {r}(\lambda )_{ac}^{bd}=(-1)^{\epsilon _a+\epsilon _b}
r(\lambda )_{ac}^{bd}.
\end{eqnarray}
For the row-to-row monodromy matrix, we also have
\begin{eqnarray}
&&\hat {r}(\lambda _1-\lambda _2 )_{a_1a_2}^{b_1b_2}
T^{(1)}(\lambda _1,\{ \mu _i\} )_{b_1}^{c_1}
T^{(1)}(\lambda _2,\{ \mu _i\} )_{b_2}^{c_2}
(-1)^{(\epsilon _{b_1}+\epsilon _{c_1})
(1+\epsilon _{b_2})}
\nonumber \\
&=&T^{(1)}(\lambda _2,\{ \mu _i\} )_{a_2}^{b_2}
T^{(1)}(\lambda ,\{ \mu _i\} )_{a_1}^{b_1}
\hat {r}(\lambda _1-\lambda _2)_{b_1b_2}^{c_1c_2}
(-1)^{(\epsilon _{a_1}+\epsilon _{b_1})(1+\epsilon _{b_2})}.
\end{eqnarray}

In order to prove that the nested monodromy matrix is indeed the
transfer matrix with reflecting boundary conditions, we need to prove
that it constitutes a commuting family. As discussed in the
last sections, we should prove that $K^{(1)}$ and ${K^{(1)}}^+$ satisfy
something like reflection equations. One can prove that
$K^{(1)}$ and ${K^{(1)}}^+$ satisfy the following graded reflection
equations in the newly defined graded sense.
\begin{eqnarray}
&&\hat {r}(\lambda -\mu )_{a_1a_2}^{b_1b_2}
K^{(1)}(\lambda )_{b_1}^{c_1}
\hat {r}(\lambda +\mu )_{b_2c_1}^{c_2d_1}                           
K^{(1)}(\mu )_{c_2}^{d_2}
(-)^{(\epsilon _{b_1}+\epsilon _{c_1})(1+\epsilon _{b_2})}
\nonumber \\
&=&K^{(1)}(\mu )_{a_2}^{b_2}\hat {r}(\lambda +\mu )_{a_1b_2}^{b_1c_2}     
K^{(1)}(\lambda )_{b_1}^{c_1}
\hat {r}(\lambda -\mu )_{c_2c_1}^{d_2d_1}                            
(-)^{(\epsilon _{b_1}+\epsilon _{c_1})(1+\epsilon _{c_2})},
\end{eqnarray}
\begin{eqnarray}
&&\hat {r}(-\lambda +\mu )_{a_1a_2}^{b_1b_2}
{K^{(1)}}^+(\lambda )_{b_1}^{c_1}
\hat {r}(-\lambda -\mu )_{b_2c_1}^{c_2d_1}                           
{K^{(1)}}^+(\mu )_{c_2}^{d_2}
(-)^{(\epsilon _{b_1}+\epsilon _{c_1})(1+\epsilon _{b_2})}
\nonumber \\
&=&{K^{(1)}}^+(\mu )_{a_2}^{b_2}\hat {r}(-\lambda -\mu )_{a_1b_2}^{b_1c_2}     
{K^{(1)}}^+(\lambda )_{b_1}^{c_1}
\hat {r}(-\lambda +\mu )_{c_2c_1}^{d_2d_1}                            
(-)^{(\epsilon _{b_1}+\epsilon _{c_1})(1+\epsilon _{c_2})}.
\end{eqnarray}
We see that the second relation is consistent with the cross-unitarity
relation $\hat {r}^{st_1}_{12}(-\lambda )
\hat {r}^{st_1}_{21}(\lambda )=-sin^2(\lambda )\cdot id.$.
Thus the nested transfer matrix is proved to constitute a commuting
family. We can still use the graded algebraic Bethe ansatz method to
find its eigenvalue and eigenvector.

\subsection{Algebraic Bethe ansatz method for BF six vertex model
with boundaries and the final results for BFF case}
Denote the double-row monodromy matrix as
\begin{eqnarray}
{\cal {T}}^{(1)}(\lambda ,\{ \mu _i\} )
&=&\left( \begin{array}{cc}
{\cal {A}}^{(1)}(\lambda ) &{\cal {B}}^{(1)}(\lambda ) \\
{\cal {C}}^{(1)}(\lambda ) &{\cal {D}}^{(1)}(\lambda ) \end{array}
\right).
\end{eqnarray}
For convenience, we need the following transformation
\begin{eqnarray}
{\cal {A}}^{(1)}(\lambda )=\tilde {\cal {A}}^{(1)}(\lambda )
-\frac {sin(\eta )e^{-2i\lambda }}{sin(2\lambda -\eta )}
{\cal {D}}^{(1)}(\lambda ).
\end{eqnarray}
Because the nested double-row monodromy matrix satisfy
the reflection equation 
\begin{eqnarray}
&&\hat {r}(\lambda -\mu )_{a_1a_2}^{b_1b_2}
{\cal {T}}^{(1)}(\lambda )_{b_1}^{c_1}
\hat {r}(\lambda +\mu )_{b_2c_1}^{c_2d_1}                           
{\cal {T}}^{(1)}(\mu )_{c_2}^{d_2}
(-)^{(\epsilon _{b_1}+\epsilon _{c_1})(1+\epsilon _{b_2})}
\nonumber \\
&=&{\cal {T}}^{(1)}(\mu )_{a_2}^{b_2}
\hat {r}(\lambda +\mu )_{a_1b_2}^{b_1c_2}
{\cal {T}}^{(1)}(\lambda )_{b_1}^{c_1}
\hat {r}(\lambda -\mu )_{c_2c_1}^{d_2d_1}                            
(-)^{(\epsilon _{b_1}+\epsilon _{c_1})(1+\epsilon _{c_2})},
\end{eqnarray}
we have the following commutation relations:
\begin{eqnarray}
{\cal {D}}^{(1)}(\lambda )
{\cal {C}}^{(1)}(\mu )
&=&\frac {sin(\lambda -\mu +\eta )sin(\lambda +\mu )}
{sin(\lambda -\mu )sin(\lambda +\mu -\eta )}
{\cal {C}}^{(1)}(\mu )
{\cal {D}}^{(1)}(\lambda )
\nonumber \\
&-&\frac {sin(2\mu )sin(\eta )e^{i(\lambda -\mu )}}
{sin(\lambda -\mu )sin(2\mu -\eta )}
{\cal {C}}^{(1)}(\lambda )
{\cal {D}}^{(1)}(\mu )
+\frac {sin(\eta )e^{i(\lambda +\mu )}}{sin(\lambda +\mu -\eta )}
{\cal {C}}^{(1)}(\lambda )
\tilde {\cal {A}}^{(1)}(\mu ),
\end{eqnarray}
\begin{eqnarray}
\tilde {\cal {A}}^{(1)}(\lambda )
{\cal {C}}^{(1)}(\mu )
&=&\frac {sin(\lambda -\mu +\eta )sin(\lambda +\mu )}
{sin(\lambda -\mu )sin(\lambda +\mu -\eta )}
{\cal {C}}^{(1)}(\mu )
\tilde {\cal {A}}^{(1)}(\lambda )
\nonumber \\
&&-\frac {sin(\eta )sin(2\lambda )
e^{-i(\lambda -\mu )}}
{sin(\lambda -\mu )sin(2\lambda -\eta )}
{\cal {C}}^{(1)}(\lambda )
\tilde {\cal {A}}^{(1)}(\mu )
\nonumber \\
&&+\frac {sin(2\mu )sin(2\lambda )sin(\eta )e^{-i(\lambda +\mu )}}
{sin(\lambda +\mu -\eta )sin(2\lambda -\eta )sin(2\mu -\eta )}
{\cal {C}}^{(1)}(\lambda )
{\cal {D}}^{(1)}(\mu ),
\end{eqnarray}
\begin{eqnarray}
{\cal {C}}^{(1)}(\lambda )
{\cal {C}}^{(1)}(\mu )
= -\frac {sin(\lambda -\mu +\eta )}{sin(\lambda -\mu -\eta )} 
{\cal {C}}^{(1)}(\mu )
{\cal {C}}^{(1)}(\lambda  ).
\end{eqnarray}
For the local vacuum state
$|0>^{(1)}=\bar {\otimes }_{k=1}^n|0>_k^{(1)}$, we have
\begin{eqnarray}
{\cal {B}}^{(1)}(\tilde {\lambda })|0>^{(1)}&=&0,
\nonumber \\
{\cal {C}}^{(1)}(\tilde {\lambda })|0>^{(1)}&\not= &0,
\nonumber \\
\tilde {\cal {A}}^{(1)}(\tilde {\lambda })|0>^{(1)}
&=&U_1(\tilde {\lambda })
\prod _{i=1}^n[sin(\tilde {\lambda }+\tilde {\mu }_i)
sin(\tilde {\lambda }-\tilde {\mu }_i)]|0>^{(1)},
\nonumber \\
{\cal {D}}^{(1)}(\tilde {\lambda })|0>^{(1)}&=&
U_2(\tilde {\lambda })\prod _{i=1}^n        
[sin(\tilde {\lambda }+\tilde {\mu }_i-\eta )
sin(\tilde {\lambda }-\tilde {\mu }_i-\eta )]
|0>^{(1)}.
\end{eqnarray}
Acting the transfer matrix
$t^{(1)}(\tilde {\lambda })=
U_1^+(\tilde {\lambda })
\tilde {\cal {A}}^{(1)}(\tilde {\lambda })
-U_2^+(\tilde {\lambda })
{\cal {D}}^{(1)}(\tilde {\lambda })$ on the ansatz of the eigenvector
${\cal {C}}(\tilde {\mu }^{(1)}_1)
{\cal {C}}(\tilde {\mu }^{(1)}_2)\cdots 
{\cal {C}}(\tilde {\mu }^{(1)}_m)|0>^{(1)}$,
we find the eigenvalue of the nested transfer matrix as follows:
\begin{eqnarray}
\Lambda ^{(1)}(\tilde {\lambda })
&=&U_1^+(\tilde {\lambda })
U_1(\tilde {\lambda })
\prod _{i=1}^n[sin(\tilde {\lambda }+\tilde {\mu }_i)
sin(\tilde {\lambda }-\tilde {\mu }_i)]
\prod _{l=1}^m \left\{
\frac {sin(\tilde {\lambda }-\tilde {\mu }^{(1)}_l+\eta )
sin(\tilde {\lambda }+\tilde {\mu }^{(1)}_l)}
{sin(\tilde {\lambda }-\tilde {\mu }^{(1)}_l)
sin(\tilde {\lambda }+\tilde {\mu }^{(1)}_l-\eta )}\right\} 
\nonumber\\ 
&&-U_2^+(\tilde {\lambda })
U_2(\tilde {\lambda })\prod _{i=1}^n
[sin(\tilde {\lambda }+\tilde {\mu }_i-\eta )
sin(\tilde {\lambda }-\tilde {\mu }_i-\eta )] 
\nonumber \\
&&\prod _{l=1}^m
\left\{ \frac {sin(\tilde {\lambda }-\tilde {\mu }^{(1)}_l+\eta )
sin(\tilde {\lambda }+\tilde {\mu }^{(1)}_l)}
{sin(\tilde {\lambda }-\tilde {\mu }^{(1)}_l)
sin(\tilde {\lambda }+\tilde {\mu }^{(1)}_l-\eta )}\right\} ,
\end{eqnarray}
where $\tilde {\mu }^{(1)}_1, \cdots ,\tilde {\mu }^{(1)}_m$
should satisfy the following Bethe ansatz equations
\begin{eqnarray}
&&\frac {U_1^+(\tilde {\mu }_j^{(1)})
U_1(\tilde {\mu }_j^{(1)})}
{U_2^+(\tilde {\mu }_j^{(1)})
U_2(\tilde {\mu }_j^{(1)})}
\prod _{i=1}^n
\frac {sin(\tilde {\mu }_j^{(1)}+\tilde {\mu }_i)
sin(\tilde {\mu }_j^{(1)}-\tilde {\mu }_i)}
{sin(\tilde {\mu }_j^{(1)}+\tilde {\mu }_i-\eta )
sin(\tilde {\mu }_j^{(1)}-\tilde {\mu }_i-\eta )} 
=1,~~~j=1, \cdots , m.
\end{eqnarray}

The eigenvalue of the transfer matrix $t(\lambda )$
with reflecting boundary condition is finally obtained as
\begin{eqnarray}
\Lambda (\lambda )
&=&-U_3^+(\lambda )U_3(\lambda )sin^{2N}(\lambda -\eta )
\prod _{i=1}^n
\frac {sin(\lambda +\mu _i)sin(\lambda -\mu _i+\eta )}
{sin(\lambda +\mu _i-\eta )sin(\lambda -\mu _i)}
\nonumber \\
&&+sin^{2N}(\lambda )
\prod _{i=1}^n
\frac {1}
{sin(\lambda -\mu _i)sin(\lambda +\mu _i-\eta )}
\Lambda ^{(1)}(\tilde {\lambda }),
\end{eqnarray}
and $\mu _1, \cdots ,\mu _m$ should satisfy the Bethe ansatz equations:

\begin{eqnarray}
&&\frac {sin(2\tilde {\mu }_j+\eta )}
{sin(2\tilde {\mu }_j-\eta )}
\prod _{i=1,\not =j}^n \left\{ \frac 
{sin(\mu _j+\mu _i)sin(\mu _j-\mu _i+\eta )}
{sin(\tilde {\mu }_j+\tilde {\mu }_i-\eta )
sin(\tilde {\mu }_j-\tilde {\mu }_i-\eta )}\right\}
\nonumber \\
&=&\frac
{sin^{2N}(\mu _j )}{sin^{2N}(\mu _j-\eta )}
\frac {U_2^+(\tilde {\mu }_j)U_2(\tilde {\mu }_j)}
{U_3^+(\mu _j)U_3(\mu _j)}
\prod _{l=1}^m
\left\{ \frac {sin(\tilde {\mu }_j-\tilde {\mu }^{(1)}_l+\eta )
sin(\tilde {\mu }_j+\tilde {\mu }^{(1)}_l)}
{sin(\tilde {\mu }_j-\tilde {\mu }^{(1)}_l)
sin(\tilde {\mu }_j+\tilde {\mu }^{(1)}_l-\eta )}\right\} ,
\nonumber \\
&&~~~~j=1,\cdots ,n,
\end{eqnarray}
where $\tilde {\mu }=\mu -{1\over 2}\eta $.

Finally, we give a summary of $U$ and $U^+$ for BFF grading.

Case I:
\begin{eqnarray}
U_1^+(\tilde {\lambda })&=&sin(\tilde {\xi }^+-\tilde {\lambda })
e^{2i\tilde {\lambda }},
\nonumber \\
U_2^+(\tilde {\lambda })&=&
\frac {sin(2\tilde {\lambda })
sin(\tilde {\xi }^+-\tilde {\lambda })
}{sin(2\tilde {\lambda }-\eta )}e^{i(2\tilde {\lambda }-\eta )},
\nonumber \\
U_3^+(\lambda )&=&sin(\xi ^++\lambda -\eta )e^{-2i\eta }.
\end{eqnarray}

Case II:
\begin{eqnarray}
U_1^+(\tilde {\lambda })&=&sin(\tilde {\xi }^+-\tilde {\lambda })
e^{2i\tilde {\lambda }},
\nonumber \\
U_2^+(\tilde {\lambda })&=&
\frac {sin(\tilde {\lambda }+\tilde {\xi }^+-\eta )
sin(2\tilde {\lambda })}
{sin(2\tilde {\lambda }-\eta )},
\nonumber \\
U_3^+(\lambda )&=&sin(\xi ^++\lambda -2\eta )e^{-i\eta }.
\end{eqnarray}

Case I:
\begin{eqnarray}
U_1(\tilde {\lambda })&=&
\frac {sin(2\tilde {\lambda }+\eta )sin(\tilde {\xi }+\tilde {\lambda })}
{sin(2\tilde {\lambda }-\eta )}e^{-i2(\tilde {\lambda }+\eta )},
\nonumber \\
U_2(\tilde {\lambda })&=&
\frac {sin(2\tilde {\lambda }+\eta )sin(\tilde {\xi }+\tilde {\lambda })}
{sin(2\tilde {\lambda })}e^{-i(2\tilde {\lambda }+\eta )},
\nonumber \\
U_3(\lambda )&=&sin(\xi -\lambda ).
\end{eqnarray}

Case II:
\begin{eqnarray}
U_1(\tilde {\lambda })&=&
\frac {sin(2\tilde {\lambda }+\eta )sin(\tilde {\xi }+\tilde {\lambda }
-\eta )}
{sin(2\tilde {\lambda }-\eta )}e^{-i(2\tilde {\lambda }+\eta )},
\nonumber \\
U_2(\tilde {\lambda })&=&
\frac {sin(2\tilde {\lambda }+\eta )sin(\tilde {\xi }-\tilde {\lambda })}
{sin(2\tilde {\lambda })}e^{-i\eta },
\nonumber \\
U_3(\lambda )&=&sin(\xi -\lambda ).
\end{eqnarray}
As before $U$ and $U^+$ are independent of each other, so there
are four combinations for $\{ U, U^+\} $ such as $\{I, I\}$,
$\{I, II\}$, $\{II, I\}$ and $\{II, II\}$.

\section{Results for FBF grading}
The last possible grading is FBF, $\epsilon _1=\epsilon _3=1,
\epsilon _2=0$. We can analyze it in the same way as the BFF
grading. Here we just present the eigenvalue,  the corresponding
Bethe ansatz equation and the boundary factors. The
eigenvalue of the transfer matrix with refecting boundary
condition is
\begin{eqnarray}
\Lambda (\lambda )
&=&-U_3^+(\lambda )U_3(\lambda )sin^{2N}(\lambda -\eta )
\prod _{i=1}^n
\frac {sin(\lambda +\mu _i)sin(\lambda -\mu _i+\eta )}
{sin(\lambda +\mu _i-\eta )sin(\lambda -\mu _i)}
\nonumber \\
&&+sin^{2N}(\lambda )
\prod _{i=1}^n
\frac {1}
{sin(\lambda -\mu _i)sin(\lambda +\mu _i-\eta )}
\Lambda ^{(1)}(\tilde {\lambda }),
\nonumber \\
\Lambda ^{(1)}(\tilde {\lambda })
&=&-U_1^+(\tilde {\lambda })
U_1(\tilde {\lambda })
\prod _{i=1}^n[sin(\tilde {\lambda }+\tilde {\mu }_i)
sin(\tilde {\lambda }-\tilde {\mu }_i)]
\prod _{l=1}^m \left\{
\frac {sin(\tilde {\lambda }-\tilde {\mu }^{(1)}_l-\eta )
sin(\tilde {\lambda }+\tilde {\mu }^{(1)}_l)}
{sin(\tilde {\lambda }-\tilde {\mu }^{(1)}_l)
sin(\tilde {\lambda }+\tilde {\mu }^{(1)}_l+\eta )}\right\} 
\nonumber\\ 
&&+U_2^+(\tilde {\lambda })
U_2(\tilde {\lambda })\prod _{i=1}^n
[sin(\tilde {\lambda }+\tilde {\mu }_i+\eta )
sin(\tilde {\lambda }-\tilde {\mu }_i+\eta )] 
\nonumber \\
&&\prod _{l=1}^m
\left\{ \frac {sin(\tilde {\lambda }-\tilde {\mu }^{(1)}_l-\eta )
sin(\tilde {\lambda }+\tilde {\mu }^{(1)}_l)}
{sin(\tilde {\lambda }-\tilde {\mu }^{(1)}_l)
sin(\tilde {\lambda }+\tilde {\mu }^{(1)}_l+\eta )}\right\} ,
\end{eqnarray}
where $\tilde {\mu }^{(1)}_1, \cdots ,\tilde {\mu }^{(1)}_m$
should satisfy the following Bethe ansatz equations
\begin{eqnarray}
&&\frac {U_1^+(\tilde {\mu }_j^{(1)})
U_1(\tilde {\mu }_j^{(1)})}
{U_2^+(\tilde {\mu }_j^{(1)})
U_2(\tilde {\mu }_j^{(1)})}
\prod _{i=1}^n
\frac {sin(\tilde {\mu }_j^{(1)}+\tilde {\mu }_i)
sin(\tilde {\mu }_j^{(1)}-\tilde {\mu }_i)}
{sin(\tilde {\mu }_j^{(1)}+\tilde {\mu }_i+\eta )
sin(\tilde {\mu }_j^{(1)}-\tilde {\mu }_i+\eta )} 
=1,~~~j=1, \cdots , m,
\end{eqnarray}
and $\tilde {\mu }_1, \cdots, \tilde {\mu }_n$ should satisfy 
\begin{eqnarray}
1=\frac
{sin^{2N}(\mu _j )}{sin^{2N}(\mu _j-\eta )}
\frac {U_2^+(\tilde {\mu }_j)U_2(\tilde {\mu }_j)}
{U_3^+(\mu _j)U_3(\mu _j)}
\prod _{l=1}^m
\left\{ \frac {sin(\tilde {\mu }_j-\tilde {\mu }^{(1)}_l-\eta )
sin(\tilde {\mu }_j+\tilde {\mu }^{(1)}_l)}
{sin(\tilde {\mu }_j-\tilde {\mu }^{(1)}_l)
sin(\tilde {\mu }_j+\tilde {\mu }^{(1)}_l+\eta )}\right\} ,
~~~~j=1,\cdots ,n.
\end{eqnarray}
The boundary factors are as follows:

Case I:
\begin{eqnarray}
U_1^+(\tilde {\lambda })&=&sin(\tilde {\xi }^+-\tilde {\lambda })
e^{2i\tilde {\lambda }},
\nonumber \\
U_2^+(\tilde {\lambda })&=&
\frac {sin(2\tilde {\lambda })
sin(\tilde {\xi }^+-\tilde {\lambda })
}{sin(2\tilde {\lambda }+\eta )}e^{i(2\tilde {\lambda }+\eta )},
\nonumber \\
U_3^+(\lambda )&=&sin(\xi ^++\lambda -\eta ).
\end{eqnarray}

Case II:
\begin{eqnarray}
U_1^+(\tilde {\lambda })&=&sin(\tilde {\xi }^+-\tilde {\lambda })
e^{2i\tilde {\lambda }},
\nonumber \\
U_2^+(\tilde {\lambda })&=&
\frac {sin(\tilde {\lambda }+\tilde {\xi }^++\eta )
sin(2\tilde {\lambda })}
{sin(2\tilde {\lambda }+\eta )},
\nonumber \\
U_3^+(\lambda )&=&sin(\xi ^++\lambda )e^{-i\eta }.
\end{eqnarray}

Case I:
\begin{eqnarray}
U_1(\tilde {\lambda })&=&
sin(\tilde {\xi }+\tilde {\lambda })e^{-i2(\tilde {\lambda }+\eta )},
\nonumber \\
U_2(\tilde {\lambda })&=&
\frac {sin(2\tilde {\lambda }+\eta )sin(\tilde {\xi }+\tilde {\lambda })}
{sin(2\tilde {\lambda })}e^{-i(2\tilde {\lambda }+\eta )},
\nonumber \\
U_3(\lambda )&=&sin(\xi -\lambda ).
\end{eqnarray}

Case II:
\begin{eqnarray}
U_1(\tilde {\lambda })&=&
sin(\tilde {\xi }+\tilde {\lambda }+\eta )
e^{-i(2\tilde {\lambda }+\eta )},
\nonumber \\
U_2(\tilde {\lambda })&=&
\frac {sin(2\tilde {\lambda }+\eta )sin(\tilde {\xi }-\tilde {\lambda })}
{sin(2\tilde {\lambda })}e^{-i\eta },
\nonumber \\
U_3(\lambda )&=&sin(\xi -\lambda ).
\end{eqnarray}

\section{Summary and discussions}
We have studied the generalized supersymmetric $t-J$ model with boundaries
in the framework of the graded quantum inverse scattering method.
The trigonometric R-matrix of the Perk-Shultz model
is changed it to the graded one. 
Solving the reflection equation and the dual reflection equation,
we obtain two types of solutions each for three
different backgrounds FFB, BFF and FBF.
The transfer
matrix is constructed from the R-matrix and the reflecting K-matrix.
The Hamiltonian is the the supersymmetric $t-J$ model with
boundary terms.
Using the graded algebraic Bethe ansatz method, we
obtain the eigenvalues of the transfer matrix for three possible
gradings. The corresponding Bethe ansatz equations are obtained.

Comparing our results with the previous results in \cite{G},
we find that the form of Bethe ansatz equations
for BFF case in section 5 are similar to the results obtained in \cite{G}.

It is important to investigate the thermodynamic limit of the
results obtained in this paper. There, we may calculate find some physical 
quantites such as free energy, surface free energy and
interfacial tension etc..
It is also important to extend the supersymmetric $t-J$ model to a more
general supersymmetric case.

Recently, the boundary impurity problems have attracted considerable
interests \cite{WDHP, BEF, BF, HPW, FLT, ZGLG}. Studying the boundary
impurities by using three different grading is interesting and
will be left for a future study.

\vskip 1truecm
\noindent {\bf Acknowlegements}:
One of the authors, H.F. is supported by the Japan Society for
the Promotion of Science. He would like to thank the hospitality
of Department of Physics, University of Tokyo and the help of 
Wadati's group.
He thank V.E.Korepin for encouragements and useful discussions.
He also would like to thank the hospitality of
S.K.Wang and X.M.Ding when he visited
Institute of Applied Mathematics,
Academia Sinica where a part of this work was done.
We thank B.Y.Hou, K.J.Shi, R.H.Yue, W.L.Yang and L.Zhao
for useful discussions.


\newpage               
\begin{thebibliography}{99}
\bibitem{A} P.W.Anderson, Science {\bf 235}(1987) 1196;
Phys. Rev. Lett. {\bf 65}, 2306(1990).
\bibitem{ZR} F.C.Zhang and T.M.Rice, Phys. Rev. {\bf B37} (1988) 3759.
\bibitem{L} C.K.Lai, J.Math.Phys.{\bf 15}(1974)167.
\bibitem{S}B.Sutherland, Phys.Rev.{\bf B12}(1975)3795.
\bibitem{S1}P.Schlottmann, Phys.Rev.{\bf B12}(1987)5177.
\bibitem{BB}P.A.Bares and G.Blatter, Phys.Rev.Lett.{\bf 64}(1990)2567.
\\
P.A.Bares, G.Blatter and M.Ogata, Phys.Rev.{\bf B44}(1991)130
\bibitem{S2}S.Sarkar, J.Phys.{\bf A24}(1991)1137;{\bf A23}(1990)L409.
\bibitem{B0}
B.Z.Bariev, J.Phys.{\bf A27}, 3381(1994); Phys.Rev.{\bf B49},
1474(1994).
\bibitem{EK} F.H.L.Essler and V.E.Korepin,
Phys. Rev {\bf B46} (1992) 9147.
\bibitem{S3}P.Schlottmann, Int.J.Mod.Phys.{\bf B11}(1997)355.
\bibitem{TF}L.A.Takhtajan, L.D.Faddeev, Russ.Math.Surv.{\bf 34}, 11 (1979);
\bibitem{KIB}V.E.Korepin, G.Izergin and N.M.Bogoliubov,
"{\it Quantum Inverse Scattering Method, Correlation Functions and
Algebraic Bethe Ansatz}" (Cambridge University Press, Cambridge, 1992).
\bibitem{FK}A.Foerster and M.Karowski, 
Nucl. Phys. {\bf B396} (1993) 611.
\bibitem{S4}E.K.Sklyanin, J.Phys.{\bf A21} (1988) 2375.
\bibitem{C} I.V.Cherednik, Theor.Math.Phys. {\bf 17}, 77 (1983); 
{\bf 61}, 911 (1984).
\bibitem{V}H.J.de Vega, Int.J.Mod.Phys.{\bf A4}(1989)2371;
\bibitem{MN} L.Mezincescu, R.I.Nepomechie, J.Phys. {\bf A24}(1991)L19;
Mod.Phys.Lett. {\bf A6} (1991) 2497-2508
\bibitem{DD}C.Destri, H.J.de Vega, Nucl.Phys.{\bf B361}(1992)361;
Nucl. Phys. {\bf B374} (1992) 692.
\bibitem{MN1}L.Mezincescu, R.I.Nepomechie, 
"{\it Quantum Field Theory, Statistical Mechanics, Quantum 
Groups and Topology"}, eds. T. Curtright et, al 
(World Scientific, Singapore), (1992)200.
\bibitem{Z}H.Q.Zhou, J.Phys.{\bf A30}(1997)711, Phys. Lett. 
{\bf A228}(1997)48.
\bibitem{SW}M.Shiroishi, M.Wadati, J. Phys. Soc. Jpn.{\bf 66}(1997)2288.
\bibitem{FK1}A.Foerster, M.Karowski, Nucl.Phys.{\bf B408}(1994)512.
\bibitem{G}A.Gonzalez-Ruiz, Nucl.Phys.B424(1994)[FS]468.
\bibitem{DG}H.de Vega, A.Gonzalez-Ruiz, Nucl.Phys.{\bf B417}(1994)553;
Mod. Phys. Lett. {\bf A9} (1994) 2207.
\bibitem{RFH}R.H.Yue, H.Fan, B.Y.Hou, Nucl. Phys. {\bf B462}(1996) 167;
\bibitem{E}F.H.L.Essler, J.Phys.{\bf A29}(1996)6183.
\bibitem{BGZZ}A.J.Bracken, X.Y.Ge, Y.Z.Zhang and H.Q.Zhou,
cond-mat/9710141, Nucl.Phys.{\bf B516} (1998) 588.
\bibitem{GZZ}M.D.Gould, Y.Z.Zhang and H.Q.Zhou,
cond-mat/9709129, Phys.Rev.{\bf B57}(199809498.
\bibitem{CDRS}E.Corrigan, P.E.Dorey, R.H.Rietdijk, R.Sasaki,
Phys.Lett.B333(1994)83;
\bibitem{FSW}P.Fendley, S.Saleur, N.P.Warner, Nucl. Phys. {\bf B430}
(1995)577; {\bf B428} (1994) 681;
\bibitem{LMSS} A.Leclair, G.Mussardo, H.Saleur, S.Skorik, 
Nucl.Phys.{\bf B453}[FS] (1995)581;
\bibitem{BZ}M.T.Batchelor, Y.K.Zhou, Phys.Rev.Lett.{\bf 76} 
(1996)2826;
\bibitem{LGWW}
C.X.Liu, G.X.Gu, S.K.Wang, K.Wu, {\it Classification of solutions
to reflection equation of two-component systems},
hep-th/9808083.
\bibitem{BP}R.E.Behrend, P.A.Pearce, J.Phys.{\bf A29} (1996)7828;
\bibitem{GZ}S.Ghoshal, A.Zamolodchikov, Int.J.Mod.Phys.{\bf A9} 
(1994) 3841;
\bibitem{JKKKM}M.Jimbo, K.Kedem, T.Kojima, H.Konno, T.Miwa, Nucl. Phys.
{\bf B441}(1995) 437 .
\bibitem{AS}H.Asakawa, M.Suzuki, Int.J.Mod.Phys.{\bf B11}, 1137(1997);
J.Phys.{\bf A29}, 225(1996);
{\bf A29}, 7811(1996); {\bf A30}, 3741(1997).
\bibitem {FHSY} H.Fan, B.Y.Hou, K.J.Shi, Z.X.Yang,
Nucl.Phys.{\bf B478} (1996) 723; 
\\ H.Fan, Nucl. Phys. {\bf B488} (1997) 409; 
\\ H.Fan, B.Y.Hou, K.J.Shi, Nucl. Phys. {\bf B496}[PM] (1997) 551 .
\bibitem{SW1}
M.Shiroishi, M.Wadati, J.Phys.Soc.Jpn.{\bf 66}, No.1 (1997) 1;
\bibitem{Y}C.N.Yang, Phys.Rev.Lett.{\bf 19} (1967) 1312;
\bibitem{B}R.J.Baxter,"{\it Exactly Solved Models in Statistical Mechanics}",
(Academic Press, London,1982).
\bibitem{FHS} H.Fan, B.Y.Hou and K.J.Shi, Nucl.Phys.{\bf B541}(1999)483.
\bibitem{PS} J.H.Perk, C.L.Shultz, Phys.Lett.{\bf A84}(1981)3759.
\bibitem{OWA}E.Olmedilla, M.Wadati, Y.Akutsu, J.Phys.Soc.Jpn
{\bf 56}(1987)2298,1340,4374.
\bibitem{FG}H.Fan,X.W.Guan, cond-mat/9711150.
\bibitem{WDHP}Y.Wang, J.H.Dai, Z.N.Hu, F.C.Pu, Phys.Rev.Lett.
{\bf 79}(1997)1901.
\bibitem{BEF}G.Bed\"urftig, F.H.L.Essler, H.Frahm,
Nucl.Phys.{\bf B489}(1997)697.                     
\bibitem{BF}G.Bed\"urftig, H.Frahm,{\it Open t-J chain with
boundary impurities}, cond-mat/9903202.
\bibitem{HPW}H.Z.Hu, F.C.Pu, W.P.Wang, J.Phys.{\bf A31}(1998)5241.
\bibitem{FLT}A.Foerster, J.Links, A.P.Tonel, {\it Algebraic
properties of an integrable t-J model with impurities},
cond-mat/9901091.
\bibitem{ZGLG}H.Q.Zhou, X.Y.Ge, J.Links, M.D.Gould, {\it Graded
reflection equation algebras and integrable Kondo impurities in the
one-dimensional t-J model}, cond-mat/9809056.
\end{thebibliography}
\end{document}